\documentclass[12pt,preprint]{aastex}

\newcommand\Romannum[1]{{\uppercase\expandafter{\romannumeral #1}}}
\newcommand\eqnion[2]{\rm #1\;{\rm\Romannum{#2}}\relax}

\usepackage[usenames]{color}  
\newif\ifredmode
\redmodefalse

\ifredmode
  \newcommand{\red}[1]{\textcolor{red}{#1}}
\else
  \newcommand{\red}[1]{#1}
\fi

\slugcomment{To be submitted to ApJ}

\shorttitle{Dwarf starburst SN Ia host @ z=1.55}
\shortauthors{Frederiksen et al.}

\begin{document}


\title{The Dwarf Starburst Host Galaxy of a Type Ia SN at $z=1.55$ from CANDELS\thanks{Based on observations made with ESO telescopes at the La Silla Paranal Observatory under program ID 086.A-0660}}


\author{
	Teddy F. Frederiksen\altaffilmark{2},
	Jens Hjorth\altaffilmark{2},
	Justyn R. Maund\altaffilmark{2,3,4},
	Steven A. Rodney\altaffilmark{5},
	Adam G. Riess\altaffilmark{5,6},
	Tomas Dahlen\altaffilmark{6},
	Bahram Mobasher\altaffilmark{7}
}
\email{teddy@dark-cosmology.dk}

%


\altaffiltext{2}{
	Dark Cosmology Centre,
	Niels Bohr Institute,
	University of Copenhagen,
	Juliane Maries Vej 30,
	2100 Copenhagen,
	Denmark
}
\altaffiltext{3}{
	Astrophysics Research Centre,
	School of Mathematics and Physics,
	Queen's University Belfast,
	University Road,
	Belfast BT7 1NN,
	Northern Ireland, UK
}
\altaffiltext{4}{Royal Society Research Fellow}
\altaffiltext{5}{
	Department of Physics and Astronomy,
	The Johns Hopkins University,
	Baltimore, MD 2121
}
\altaffiltext{6}{
	Space Telescope Science Institute,
	Baltimore, MD 21218
}
\altaffiltext{7}{
	Department of Physics and Astronomy,
	University of California,
	Riverside, CA 9252
}


\begin{abstract}
We present VLT/X-shooter observations of a high redshift, type Ia supernova host galaxy, discovered with {\it HST}/WFC3 as part of the CANDELS Supernova project.
The galaxy exhibits strong emission lines of Ly$\alpha$, [\ion{O}{2}], H$\beta$, [\ion{O}{3}], and H$\alpha$ at $z=1.54992^{+0.00008}_{-0.00004}$.
From the emission-line fluxes and SED fitting of broad-band photometry we rule out AGN activity and characterize the host galaxy as a young, low mass, metal poor, starburst galaxy with low intrinsic extinction and high Ly$\alpha$ escape fraction.
The host galaxy stands out in terms of the star formation, stellar mass, and metallicity compared to its lower redshift counterparts, mainly because of its high specific star-formation rate.
If valid for a larger sample of high-redshift SN Ia host galaxies, such changes in the host galaxy properties with redshift are of interest because of the potential impact on the use of SN Ia as standard candles in cosmology.
\end{abstract}





\section{INTRODUCTION}
Type Ia Supernovae (SNe Ia) are cornerstones of modern cosmology because of their properties as luminous standard candles. The development of these important cosmological tools began in the late 1930's when \citet{1938PhRv...53.1019Z} and \citet{1939ApJ....90..634W} first suggested that SNe could be used as distance indicators. Theoretical developments in the 1960's suggested that SNe of type Ia form a homogenous class of objects with a measured peak magnitude of $M_B \approx -19.3 + 5\log h_{70}$ \citep[for a modern review, see][]{2000ARA&A..38..191H,2010deot.book..151K}. To first approximation, the light curves of SN Ia form a one-parameter family of models, driven by the decay of radioactive ${}^{56}\rm{Ni} \to {}^{56}\rm{Co} \to {}^{56}\rm{Fe}$. The amount of radioactive nickel produced in the initial explosion therefore dictates the shape of the light curve. Later observational work showed that the scatter in the peak magnitude is correlated with other SN properties, such as light curve shape and color \citep{1993ApJ...413L.105P,1996ApJ...473...88R,1999AJ....118.1766P}.

SN cosmology achieved its modern prominence at the close of the millennium with the discovery of the accelerating expansion of the universe, based on just a few dozen objects \citep{1999ApJ...517..565P,1998AJ....116.1009R}.  Nearly 15 years later, modern SNIa samples can now include over 500 well-studied SNe with a dispersion in peak magnitudes of $\sim$0.16 magnitudes \citep[e.g.,][]{2011ApJS..192....1C,2012ApJ...746...85S}.  At this precision, a larger sample size alone will not improve cosmological constraints, so the limiting factor is our understanding of systematic effects.

Among the major concerns for systematic biases is the fact that we still do not have a complete or conclusive description of the SN Ia progenitor systems.  Indeed, there may be several viable progenitor pathways \citep[known as single and double degenerate models, see][for a review]{2012NewAR..56..122W}, possibly leading to slightly different explosion characteristics.  One might expect different progenitor pathways to be correlated with differences in SN environment, and such correlations with host galaxy properties have recently been observed \citep{2000AJ....120.1479H,2010ApJ...715..743K,2010ApJ...722..566L,2010MNRAS.406..782S}. Correcting for this effect with measurement of the host galaxy stellar mass brings the dispersion in absolute peak magnitude down to $\sim0.1$ magnitudes \citep{2011ApJS..192....1C,2011MNRAS.418.2308M}. This signature of environmental effects calls for further characterization of the host galaxies when SNe are used for cosmography.
\citet{2006ApJ...648..884R} discuss how a change in the progenitor population (like progenitor metallicity and age) at $1.5<z<3.0$ could affect the inferred distance in a way inconsistent with dark energy models. The redshift window $1.5<z<3.0$ is therefore favorable for disentangling systematic effects arising from environment.

The Cosmic Assembly Near-IR Deep Extragalactic Legacy Survey (CANDELS) survey (Grogin et al. 2011) is a Hubble Space Telescope (HST) multi-cycle treasury (MCT) program designed to detect high redshift SNe. The CANDELS collaboration is surveying five well-observed fields (GOODS-N, GOODS-S, COSMOS, EGS and UDF). With this observation strategy CANDELS will find SNe Ia out to redshifts of $\sim2$ \citep[see][]{2012ApJ...746....5R}. The first SN detected in the CANDELS survey was discovered 2010 October 14 in the GOODS-S field and was nicknamed SN Primo. \citet{2012ApJ...746....5R} present the light curve and grism spectrum of this SN, concluding that SN Primo was of type Ia.

The aim of this paper is to characterize the host galaxy of SN Primo. 
We  derive its spectral properties from spectroscopic emission-line fluxes and fit the spectral energy distribution (SED) based on broad-band photometry to constrain its stellar population.
We then compare the properties of the host galaxy of SN Primo with its counterparts at lower redshifts and discuss sources of bias when using high redshift SNe as standard candles.

The paper is structured as follows: In Section \ref{sec_obs} we present the spectra and photometric data. In Section \ref{sec_analysis} we perform the SED fitting using broad-band photometry to get stellar mass and stellar age and we derive the emission-line fluxes and calculate spectral properties like metallicity and the Lyman-$\alpha$ escape fraction. Finally, a discussion  and conclusions are given in Section \ref{sec_discus}.
We assume a flat $\Lambda$CDM cosmology, with $H_0=70.2$ km s$^{-1}$ Mpc$^{-1}$ and $\Omega_m=0.274$ \citep{2011ApJS..192...18K}. All magnitudes given in this paper are AB magnitudes.

\section{DATA}\label{sec_obs}
\begin{figure}
	\epsscale{0.5}
	\plotone{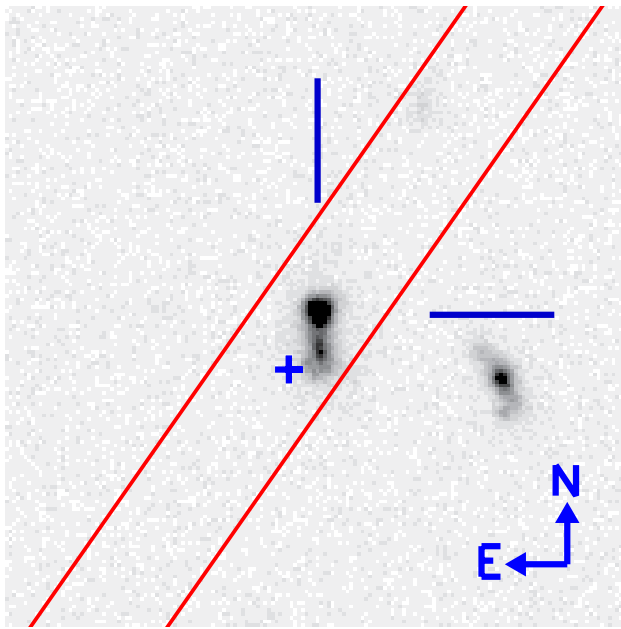}
	\caption{Pre-explosion {\it HST} F775W (i-band, rest frame u-band) image with the X-shooter slit configuration overlaid (red). The peak luminosity of the host galaxy is marked by ticks (blue). The size of the tick marks is 1\arcsec{}, corresponding to 8.6 kpc at the redshift of SN Primo. The faint part to the south is also part of the galaxy \citep[see][]{2012ApJ...746....5R}. The location of the SN is marked by a plus (blue). 
	\label{fig_slitpos}}
\end{figure}
\subsection{Spectroscopic Data}
The host of SN Primo is located in the GOODS-S field at R.A.~=~$03^{\rm h}32^{\rm m}22\fs64$ and decl.~=~$-27^\circ$46$\arcmin$38$\farcs$66 (J2000).
The spectroscopic observations were performed on 2010 October 16, using the X-shooter instrument on the ESO Very Large Telescope (VLT) at Paranal Observatory, Chile \citep{2006SPIE.6269E..98D,2011A&A...536A.105V}.
X-shooter is a cross-dispersed Echelle spectrograph with a large wavelength coverage from the UV to the Near IR (300 -- 2500 nm). This is achieved by splitting the light beam into three wavelength regions and sending them into three different spectrographs (so-called arms) designated UVB (for {UV and Blue}, $\sim$ 300 -- 550 nm), VIS (for {Visual}, $\sim$ 550 -- 1000 nm), and NIR (for {Near IR}, $\sim$ 1000 -- 2500 nm).
For the observation we used an ABAB on-source nodding template with an exposure time of 1.3 hr (4$\times$1200 sec). A $0\farcs9$ slit\footnote{A $1\farcs0$ slit in the UVB arm.} with a PA of $-35^\circ$ E of N was placed to cover both the host and the SN (see Figure \ref{fig_slitpos}). The observations were conducted under photometric conditions, with a median seeing\footnote{As measured by the Paranal on-site seeing monitor} of $0\farcs 54$.

The X-shooter spectra were reduced using the official X-shooter pipeline\footnote{See, \url{http://www.eso.org/sci/software/pipelines/}} v1.3.7.
We achieved resolving powers of $R=5400 \pm 360$ (UVB), $R=7450\pm300$ (VIS), and $R=5800\pm180$ (NIR).
The extraction of the object spectra was conducted with our own IDL script, and flux calibration was done using the {\it HST} flux standard star, GD71\footnote{See, \url{http://www.eso.org/sci/observing/tools/standards/spectra/gd71.html}}.

Even though the SN was still bright it was not detected in the spectrum (see Figure \ref{fig_spec}); only narrow emission lines from the host galaxy were visible with no trace of any continuum emission from either host or SN.

\subsection{Flux Calibration Quality}\label{sec_fluxcal}
We checked the quality of the flux calibration using a telluric standard star (HIP~018926) taken prior to the science exposure.
The telluric standard star was of stellar type B3V, and has a high signal-to-noise ratio ($\sim50 - 300$ per wavelength bin, $0.04-0.1$~nm per bin). The telluric standard star was reduced, extracted and flux calibrated in the same way as the science exposure. The flux calibrated spectrum agrees with the published photometric data points. A model spectrum of a B3V star was scaled to the photometric data and compared to the flux calibrated spectrum. The agreement between the model and the spectrum was within 5--10\%, consistent with the expected quality of the flux calibration for X-shooter\footnote{\url{http://www.eso.org/sci/facilities/paranal/instruments/xshooter/doc/}} at the wavelengths of the lines detected in the science exposures (except at the location of Ly$\alpha$). At an observed wavelength of 310~nm Ly$\alpha$ is close to the spectral lower limit of the UVB arm (300~nm).
\red{The discrepancy between the model and the telluric spectrum is $\sim40\%$, i.e., the conversion factor between counts and cgs units is too large and will overestimate the flux of Ly$\alpha$. We are therefore cautious when drawing conclusions based on the flux of the Ly$\alpha$ line.}

\subsection{Photometric Data}
\begin{figure}
	\epsscale{1.0}
	\plotone{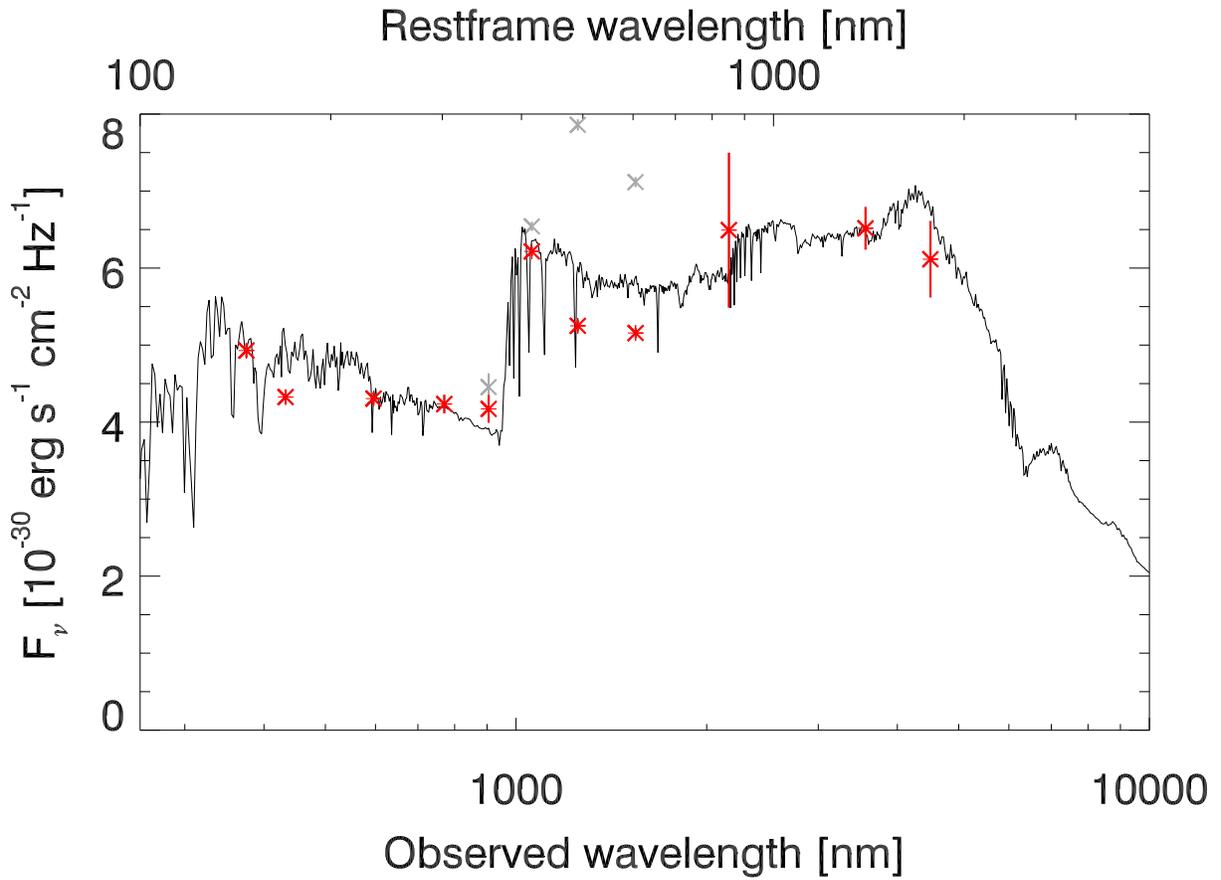}
	\caption{The SED of the SN Primo host derived from broad-band photometry.
	In red (gray) are the photometric points after (before) the subtraction of the emission-line flux.
	Overplotted is the best-fitting model SED from FAST.\label{fig_sed}}
\end{figure}
To construct the SED for the host of SN Primo, we use photometry from the F160W filter (H-band) selected TFIT catalogue. The photometry in each band is carried out using the TFIT algorithm \citep{2007PASP..119.1325L}. This method performs point-spread function (PSF) matched photometry uniformly across different instruments and filters, despite their large variations in PSFs and pixel scales. The final catalogue has photometry in VLT/VIMOS (U-band), {\it HST}/ACS (F435W, F606W, F775W, and F850LP), {\it HST}/WFC3 (F105W, F125W, and F160W), VLT/ISAAC (K$_s$), and two {\it Spitzer}/IRAC channels (3.6~$\mu$m and 4.5~$\mu$m).
\red{The SN search uses the {\it HST}/WFC3 bands. To get SN free photometry in these bands a set of pre-explosion images from another {\it HST}/WFC3 survey \citep[GO-11563, PI:Illingworth, see e.g.][]{2010ApJ...709L..16O} was used.}
The photometry is listed in Table 1.
The {\it HST} (WFC3)  observations are performed as a part of CANDELS project and are further described in \citet{2011ApJS..197...35G} and \citet{2011ApJS..197...36K}. More details on the rest of the filters and observations are given in \citet{2010ApJ...724..425D}.

\section{ANALYSIS}\label{sec_analysis}

\subsection{Broad-Band SED Fitting}\label{sec_sed}
The broad-band SED of the host of SN Primo (Table \ref{tab_photometry}) covers rest frame UV to near-IR (200 -- 2000 nm). We use the SED fitting code FAST \citep{2009ApJ...700..221K} to derive properties such as stellar mass, $M_*$, and stellar age, $t_*$.
The photometric data is corrected for the strong emission lines detected in the spectrum by subtracting the line flux from the corresponding filters.
The strong emission lines are not modeled by the stellar population synthesis models of \citet{2003MNRAS.344.1000B} used by FAST. For each filter that has strong emission lines the flux is corrected as
\begin{equation}
	F_\nu^{\rm (corr)}=F_{\nu}-\frac{F_{\rm line}}{\Delta\nu} ,
\end{equation}
where $F_{\nu}$ is the flux density in the broad-band filter, $F_{\rm line}$ is the flux in the emission line, and \red{$\Delta\nu=\int T(\nu)d\nu \cdot T(\nu_{\rm line})^{-1}$ is the integral of the filter curve, corrected for the transmission of the filter at the location of the emission line. We use the same transmission function that is used in FAST for each filter.}

The emission-line-subtracted SED is corrected for Galactic extinction \citep[$E(B-V)=0.008$\footnote{Quoted from the NASA/IPAC Extragalactic Database (NED) website: \url{http://ned.ipac.caltech.edu/}}]{1998ApJ...500..525S} using a Galactic extinction-law \citep{1989ApJ...345..245C} and $R_V=3.1$. We use FAST to fit the corrected SED assuming a \citet{2003PASP..115..763C} initial mass function (IMF) and three different star-formation histories (SFH, see Table \ref{tab_sed} and Figure \ref{fig_sed}).
The masses, ages and star-formation rates (SFR) derived in Table \ref{tab_sed} assuming different SFH agree within the $1\sigma$ uncertainties.

\red{
We also run the SED fitting without any correction to the broad-band SED, but excluding the J and H band to get a second measure of the physical parameters. This second measure quantifies the systematic shift that the correction procedure can put on the physical parameters. We check that the best fitting parameters of this second fit is within the $1\sigma$ error bars. For reference the shift in $\log(M_*)$ is $0.07$ higher for the second fit, compared to the upper error bar of $0.13$ on our main SED fit.
}

\subsection{Resampling of the X-shooter Spectrum}\label{sec_resampling}
\begin{figure}
	\epsscale{1.0}
	\plotone{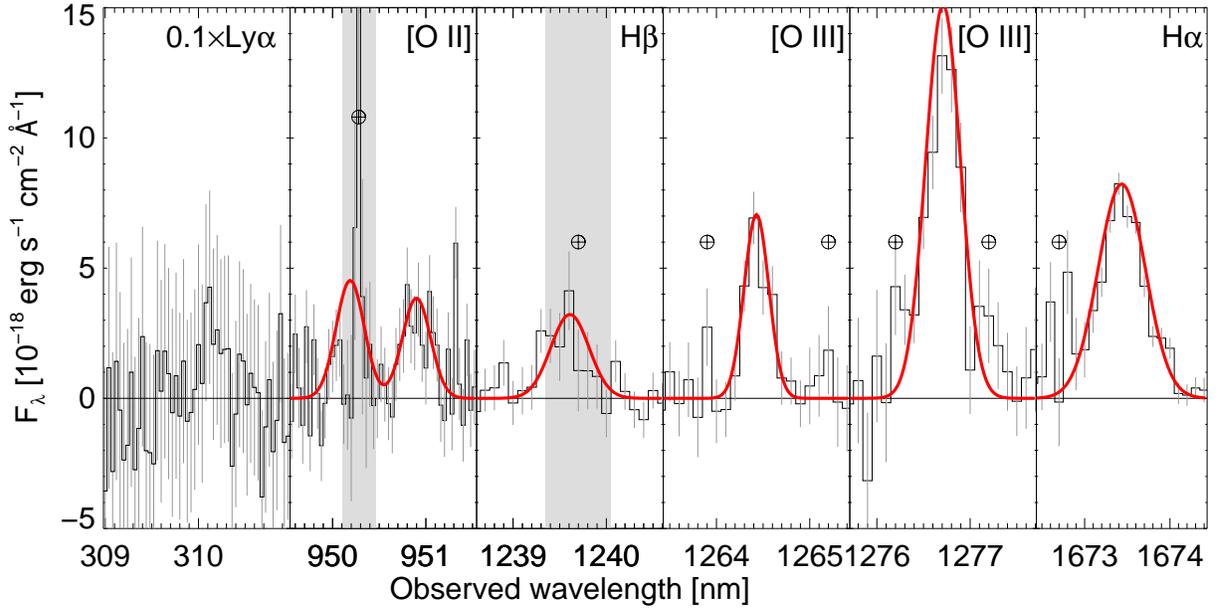}
	\caption{The detected emission lines in the X-shooter spectrum of the host of SN Primo. The spectrum is flux-calibrated and corrected for Galactic extinction. The solid (red) line shows the best fitting gaussian line profiles. The [\ion{O}{2}] ${\lambda\lambda3726,3729}$ line is fitted with a double gaussian line profile. The gray bands mark regions excluded due to sky lines. In the H$\beta$ fit the line center was fixed to $\lambda_{\rm H\beta}(1+z)$, where $\lambda_{\rm H\beta}=486.1325$ {nm} and $z$ is the redshift.
	Ly$\alpha$ is detected, but located close to the Earth's atmospheric UV cutoff. 
	\label{fig_spec}}
\end{figure}

\red{We correct the spectrum for Galactic extinction in the same manner as the broad-band photometry.}
To obtain a robust estimate of the uncertainties in the spectral quantities such as the metallicity or line ratios, we re-sample the X-shooter spectrum 10\,000 times. For each wavelength bin we resample the flux using the error spectrum (assuming gaussian error). In each iteration the spectral lines are fitted with a gaussian line profile and the centroid, the Full Width at Half Maximum (FWHM), and the total flux is calculated (see Table \ref{tab_lines} and Figure \ref{fig_spec}).
The redshift is determined from H$\alpha$, [\ion{O}{2}] ${\lambda3729}$, [\ion{O}{3}] ${\lambda\lambda4959,5007}$ in each iteration.
All reported values that are derived from the spectrum are the median values and 68\% error bars of the 10\,000 samplings.
The heliocentric velocity correction is 6.54 km~s$^{-1}$, calculated using the IRAF task {\tt rvcorrect}.

Special care is taken when fitting [\ion{O}{2}] ${\lambda\lambda3726,3729}$ and H$\beta$ in each resampling:
The blue component of the [\ion{O}{2}] doublet, [\ion{O}{2}] ${\lambda3726}$, is located on top of a sky line. After masking out the sky line it is impossible to fit the peak of [\ion{O}{2}] ${\lambda3726}$. We therefore fit a double-gaussian line-profile to [\ion{O}{2}] ${\lambda\lambda3726,3729}$. We fix the peak of the blue components, $\lambda_{\rm blue}$, to the peak of the red component, $\lambda_{\rm red}$, by requiring $\lambda_{\rm blue}/\lambda_{\rm red}=372.6032 \rm{nm}/372.8815 \rm{nm}$. The flux ratio of the two components is left as a free parameter.

H$\beta$ is also located on top of a sky line with the wings visible. We remove the sky line in the same manner as for [\ion{O}{2}] ${\lambda3726}$ and fix the wavelength, $\lambda_0$, and FWHM of the fit. $\lambda_0$ is fixed to $\lambda_{\rm H\beta}(1+z)$, where $\lambda_{\rm H\beta}=486.1325 \rm{nm}$.
The FWHM is fixed to the measured FWHM of H$\alpha$ in velocity units. The instrumental broadening of spectral lines is constant if measured in velocity units and therefore affects H$\alpha$ and H$\beta$ equally.
The derived flux may be biased if a gaussian line profile is not a correct description of the line.
Due to the uncertainties in the H$\beta$ detection we will not use the derived flux, other than for constraining the Balmer decrement. For all other purposes we set the flux of H$\beta$ equal to the flux of H$\alpha$ divided by 2.86 (see discussion in Section \ref{sec_extinction}).

We do not detect [\ion{N}{2}] ${\lambda6583}$ in the spectrum. To derive an upper limit of the flux, we measure the standard deviation of the flux density at the location of the line, $\lambda_{\rm NII}(1+z) \pm 2\Delta\lambda$, where $\lambda_{\rm NII}=658.346 \rm{nm}$ and $\Delta\lambda=\lambda/R$ is the size of one resolution element. Table \ref{tab_lines} lists the $5\sigma$ upper limit of the non-detection.

\subsection{Host Extinction}\label{sec_extinction}
We correct the X-shooter spectrum for Galactic extinction in the same manner as for the broad-band SED.
We test if the $A_V$ from SED fitting is consistent with the spectrum.
To gauge the intrinsic extinction from the spectrum we measure the Balmer decrement, $\rm H\alpha / \rm H\beta$. By comparing the measured Balmer decrement, $B$, with the expected $B_0=2.86$ given in \citet{2006agna.book.....O} (case B recombination, $T_e=10^4$ K), we calculate the extinction as
\begin{eqnarray}
	A_V = -2.5 \log \left( \frac{B}{B_0} \right) \frac{k(V)}{k(H\alpha) - k(H\beta)} ,
\end{eqnarray}
where $k(\lambda)=A_\lambda/E(B-V)$: $k(V)\equiv R_V=3.1$, $k(H\alpha)=2.468$, and $k(H\beta)=3.631$ \citep{2001PASP..113.1449C}. We assume $R_V=3.1$ because the SED and the spectrum probe the luminosity weighted average $R_V$ of the host of SN Primo and not just the SN sight line, where a lower $R_V$ (down to $\sim1.7$) can be measured \citep{2011arXiv1111.4463P}.

The value $A_V =0.6^{+1.1}_{-0.7}$ derived from the Balmer decrement is consistent with the value derived from the SED fitting. The large uncertainty in $A_V$ is due to the difficulty in estimating the H$\beta$ flux (see Section \ref{sec_resampling}).
\red{For reference, the extinction derived from the SN light curve is $A_V=0.14\pm0.14$ \citep{2012ApJ...746....5R}, but  it does not have to be linked to the (luminosity weighted) average of the galaxy as a whole.}

\subsection{Metallicity}
\begin{figure}
	\epsscale{0.492}
	\plotone{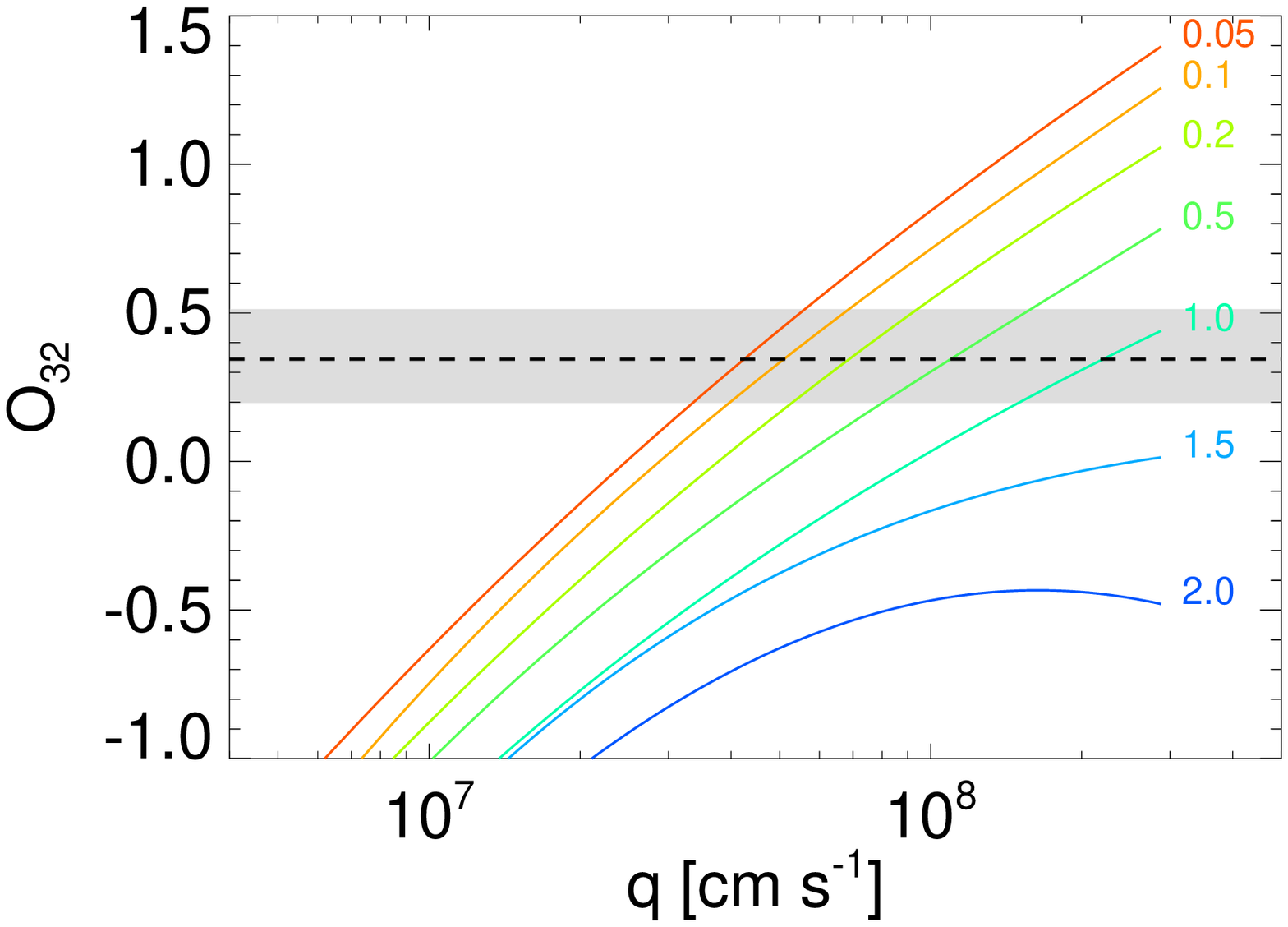}
	\plotone{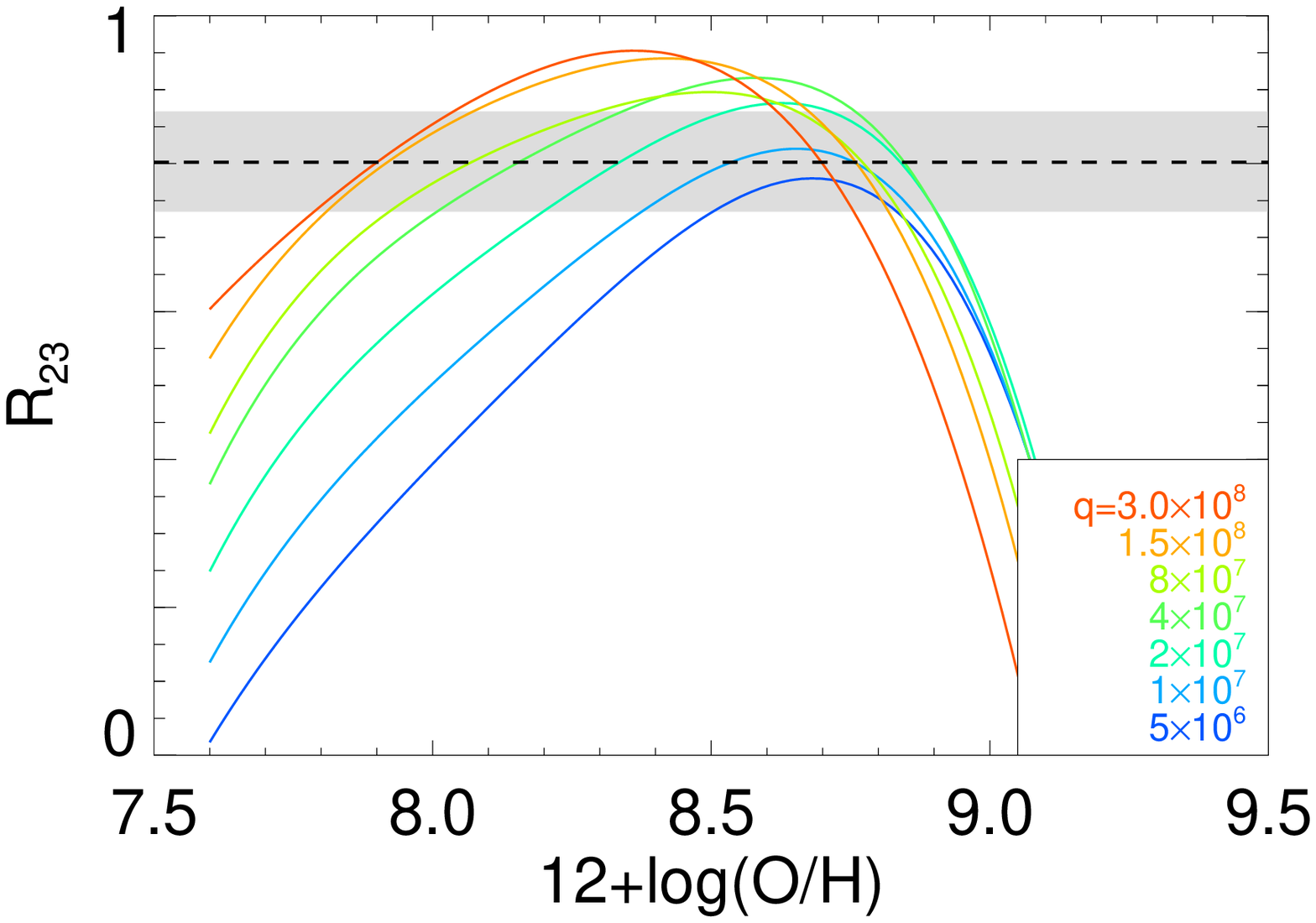}
	\caption{The $O_{32}$ and $R_{23}$ line ratios of the host of SN Primo together with the photo-ionization models of \citet{2002ApJS..142...35K}.
	The dashed lines and gray bands show the line ratios and 1$\sigma$ error bars derived from the spectrum.
	Left: $O_{32}$ (Equation \ref{eqn_O32}) versus the ionization parameter $q$.
	 From top down the metallicities are $Z$ = 0.05, 0.1, 0.2, 0.5, 1.0, 1.5, and 2.0 $Z_\odot$.
	Right: $R_{23}$ (Equation \ref{eqn_R23}) versus the metallicity.
	From the top down $q$ = 30, 15, 8, 4, 2, 1, 0.5 $\times$ $10^7$ cm s$^{-1}$. 
	\label{fig_kewley}}
\end{figure}
Given that we do not detect [\ion{N}{2}] or [\ion{S}{2}] lines in our spectrum we will use the line ratio,
\begin{equation}\label{eqn_R23}
	R_{23} = \log\left(
		\frac{
			[\eqnion{O}{2}]_{\lambda\lambda3726,3729}
			+[\eqnion{O}{3}]_{\lambda\lambda4959,5007}
		}{H\beta}
	\right),
\end{equation}
to determine the metallicity.
We take the average of the two $R_{23}$ calibrations \citep{1991ApJ...380..140M,2004ApJ...617..240K} as used in \citet{2008ApJ...681.1183K} whose procedure we follow.

The $R_{23}$ diagnostic has the problem of being double valued, meaning that from a measured $R_{23}$ value two metallicities can be inferred (see Figure \ref{fig_kewley}). We therefore need an independent measure to break this degeneracy. The upper limit on [\ion{N}{2}] ${\lambda6586}$ gives an upper limit on
\begin{equation}
	\log \left(
		\frac{[\eqnion{N}{2}]_{\lambda6586}}{[\eqnion{O}{2}]_{\lambda\lambda3726,3729}}
	\right) < -1.0.
\end{equation}
This constrains the metallicity to the low metallicity branch of $R_{23}$ \citep[see][their Figure 3]{2002ApJS..142...35K}. The low metallicity branch of $R_{23}$ changes with ionization parameter, $q$. To break the $q$-degeneracy, we need the line ratio
\begin{equation}\label{eqn_O32}
	O_{32}=\log\left(
		\frac{[\eqnion{O}{3}]_{\lambda 5007}}{[\eqnion{O}{2}]_{\lambda\lambda3726,3729}}
	\right),
\end{equation}
see Figure \ref{fig_kewley}.
The procedure of \citet{2004ApJ...617..240K} is to iterate back and forth between the two plots of Figure \ref{fig_kewley} until the estimates of metallicity and $q$ converge. In \citet{2004ApJ...617..240K} convergence is achieved after three iterations. We use 10 iterations in our implementation as this makes the convergence independent of the choice of initial guess. The metallicities derived from the two techniques are within the 0.1~dex of each other, which is the expected scatter of the two technics \citep{2002ApJS..142...35K,2008ApJ...681.1183K}. The metallicity of the SN Primo host is $12+\log(\frac{O}{H}) = 8.12^{+0.09}_{-0.10}$ or $Z=0.27\pm0.06 Z_\odot$, assuming a solar abundance of $12+\log(\frac{O}{H}) = 8.69$ \citep{2009ARA&A..47..481A}.

\subsection{Star formation}
\begin{figure}
	\epsscale{1.0}
	\plotone{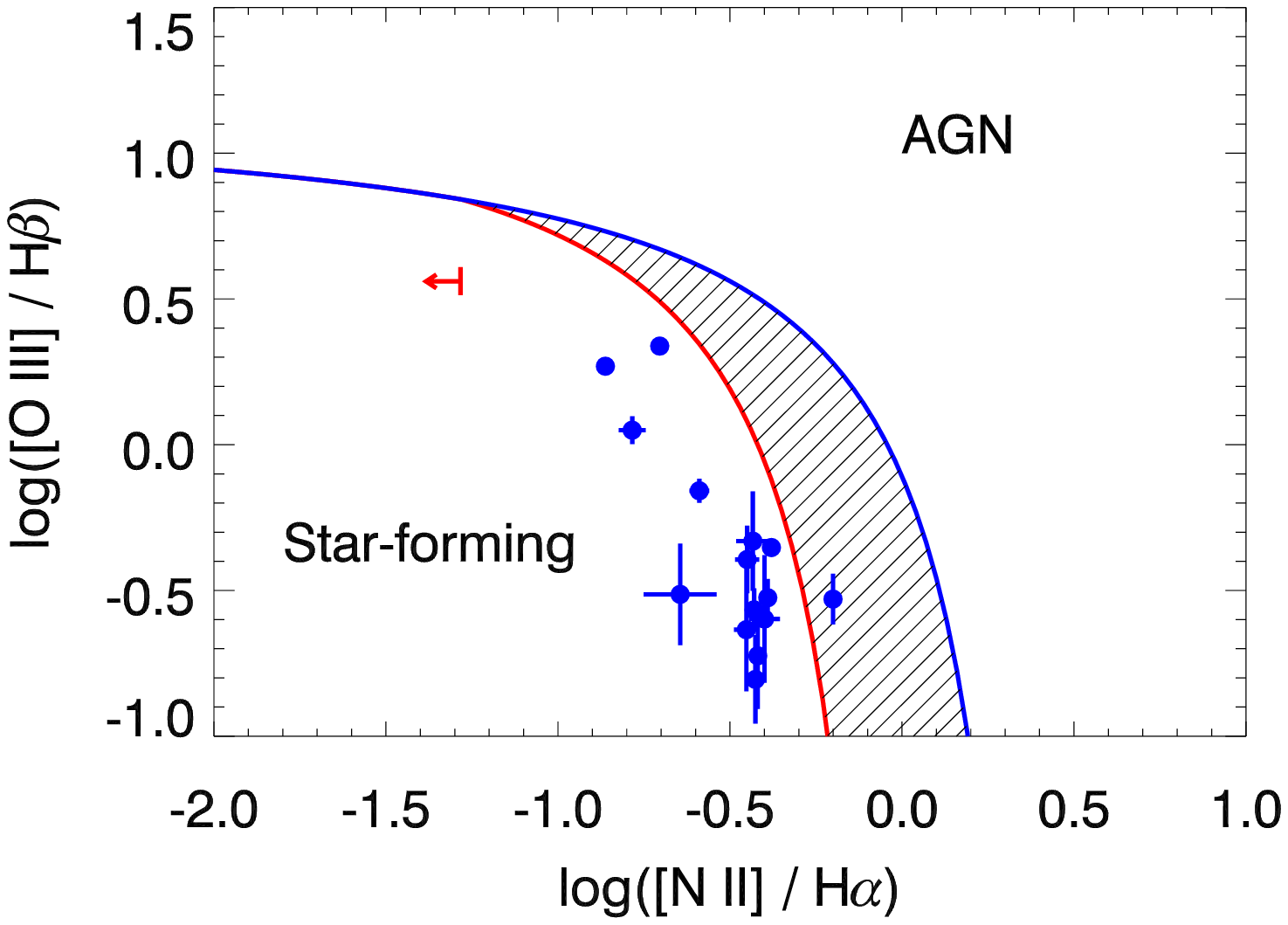}
	\caption{BPT diagnostic diagram \citep*{1981PASP...93....5B}. The x-axis denotes the ratio log([\ion{N}{2}] ${\lambda6586}$/H$\alpha$) and the y-axis the ratio log([\ion{O}{3}] ${\lambda5007}$/H$\beta$). The data point marks the host galaxy of SN Primo, with the arrow denoting the upper-limit of the ratio derived from the non-detection of [\ion{N}{2}] ${\lambda6586}$. The vertical bar denotes the 1$\sigma$ error bar of the ratio. \red{For illustrative purposes we overplot the 15 emission-line hosts from the \citet{2010ApJ...722..566L} sample of SN Ia host galaxies with SDSS spectra.}
	\label{fig_BPT}}
\end{figure}
We check whether the emission lines of the host of SN Primo are powered by star formation or AGN activity by plotting log([\ion{N}{2}] ${\lambda6586}$/H$\alpha$) versus log([\ion{O}{3}] ${\lambda5007}$/H$\beta$) in a BPT diagnostics diagram \citep*{1981PASP...93....5B}. The host of SN Primo is located in the star-forming region of Figure \ref{fig_BPT}. We therefore conclude that the H$\alpha$ flux is powered by star formation. We derive the SFR from the H$\alpha$ luminosity. We report the SFR for different IMFs for comparison \citep{1998ARA&A..36..189K,2004MNRAS.351.1151B,2010MNRAS.408.2115M}.
Using the stellar mass from SED fitting we calculate the specific SFR, sSFR~=~SFR/$M_*$ from the spectrum and obtain a value of $\sim 10^{-8}$ yr$^{-1}$, independent of the IMF and SFH, making the host of SN Primo a starburst galaxy.
\red{Our definition of a starburst is based on the sSFR \citep[see, e.g.][for a review]{2006ApJ...648..868S}, see Section \ref{sec_discus} for discussion on other definitions.}

\subsection{Lyman-$\alpha$}\label{sec_LyA}
\begin{figure}
	\epsscale{1.0}
	\plotone{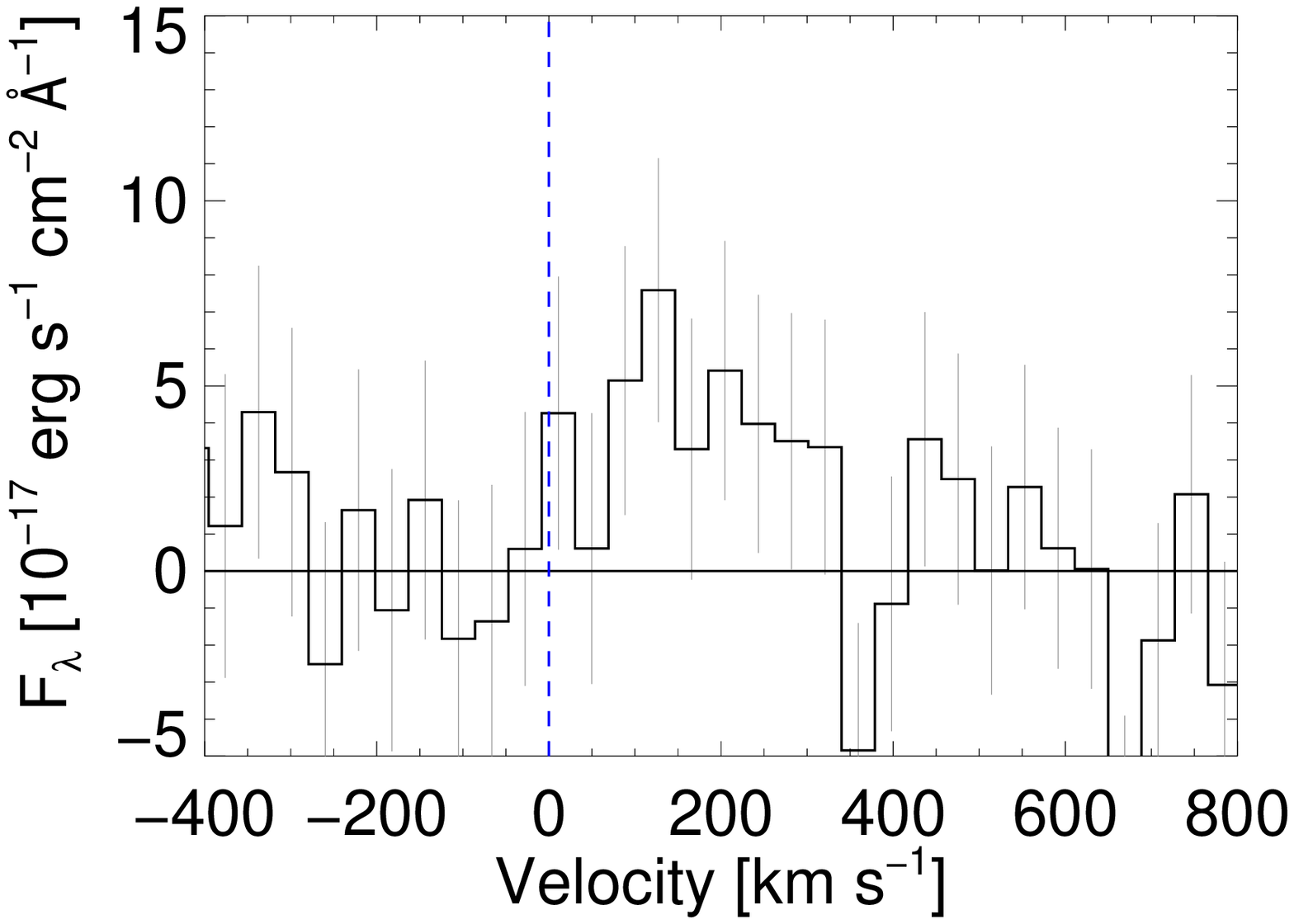}
	\caption{The Lyman-$\alpha$ line in the SN Primo host spectrum.
	The systemic velocity is determined from H$\alpha$, [\ion{O}{3}] ${\lambda\lambda4959,5007}$, and [\ion{O}{2}] ${\lambda3729}$.
	\label{fig_LyA}}
\end{figure}
\red{We detect Ly$\alpha$ emission at $2.8\sigma$ in the spectrum (see Figure \ref{fig_LyA}). This is possibly the lowest redshift ground-based detection of a cosmological Ly$\alpha$ emitter. The significance of the detection is independent of the systematic error in conversion factor between counts and cgs units (i.e. the flux calibration) at Ly$\alpha$ (Section \ref{sec_fluxcal}).}

\red{Given the low significance of the detection we can only give an order-of-magnitude estimate of the Ly$\alpha$ escape fraction \citep[as defined in][among others]{2009A&A...506L...1A,2011ApJ...730....8H},
\begin{eqnarray}\label{egn_fesc}
	f_{esc}=\frac{F_{\rm Ly\alpha}}{8.7 F_{\rm H\alpha}}.
\end{eqnarray}
We derive the line flux $F_{\rm Ly\alpha}$ by co-adding the flux in all pixels from $\lambda=309.97-310.68$ nm (corresponding to $v=0 - 600$ km s$^{-1}$). The derived flux estimate is corrected for extinction in the host galaxy.
The deviation from the expected value of 8.7 \citep[case B recombination,][]{1971MNRAS.153..471B} will be due to conditions in the interstellar medium (ISM) like the presence of dust, ISM clumpiness or due to geometric effects that will suppress or enhance the amount of Ly$\alpha$ photons that can escape the galaxy. At a redshift of $z=1.5$ the universe is fully ionized, absorption of Ly$\alpha$ in the intergalactic medium is therefore not important.
We will not try to distinguish between these different scenarios. We include a systematic uncertainty of $40\%$ in the derived Ly$\alpha$ escape fraction, due to the uncertainty in the conversion factor between counts and cgs units at the Ly$\alpha$ wavelength.}




\section{DISCUSSION \& CONCLUSIONS}
We have performed a photometric and spectroscopic study of the SN Primo host galaxy.
We find a young Large Magellanic Cloud (LMC) sized ($\sim4.5$~kpc) galaxy with LMC-like ($\sim\frac{1}{3}Z_\odot$) metallicity and low intrinsic extinction.
We confirm that the emission lines are generated by star formation and derive a SFR of almost one order of magnitude larger than that of the LMC.
The stellar mass derived from SED fitting is one order of magnitude lower then the LMC.
From the Ly$\alpha$ line we estimate a high escape fraction of Ly$\alpha$ 
photons.
All  host properties are summarized in Tables \ref{tab_sed} and \ref{tab_results}.

\label{sec_discus}
\begin{figure}
	\epsscale{1.0}
	\plotone{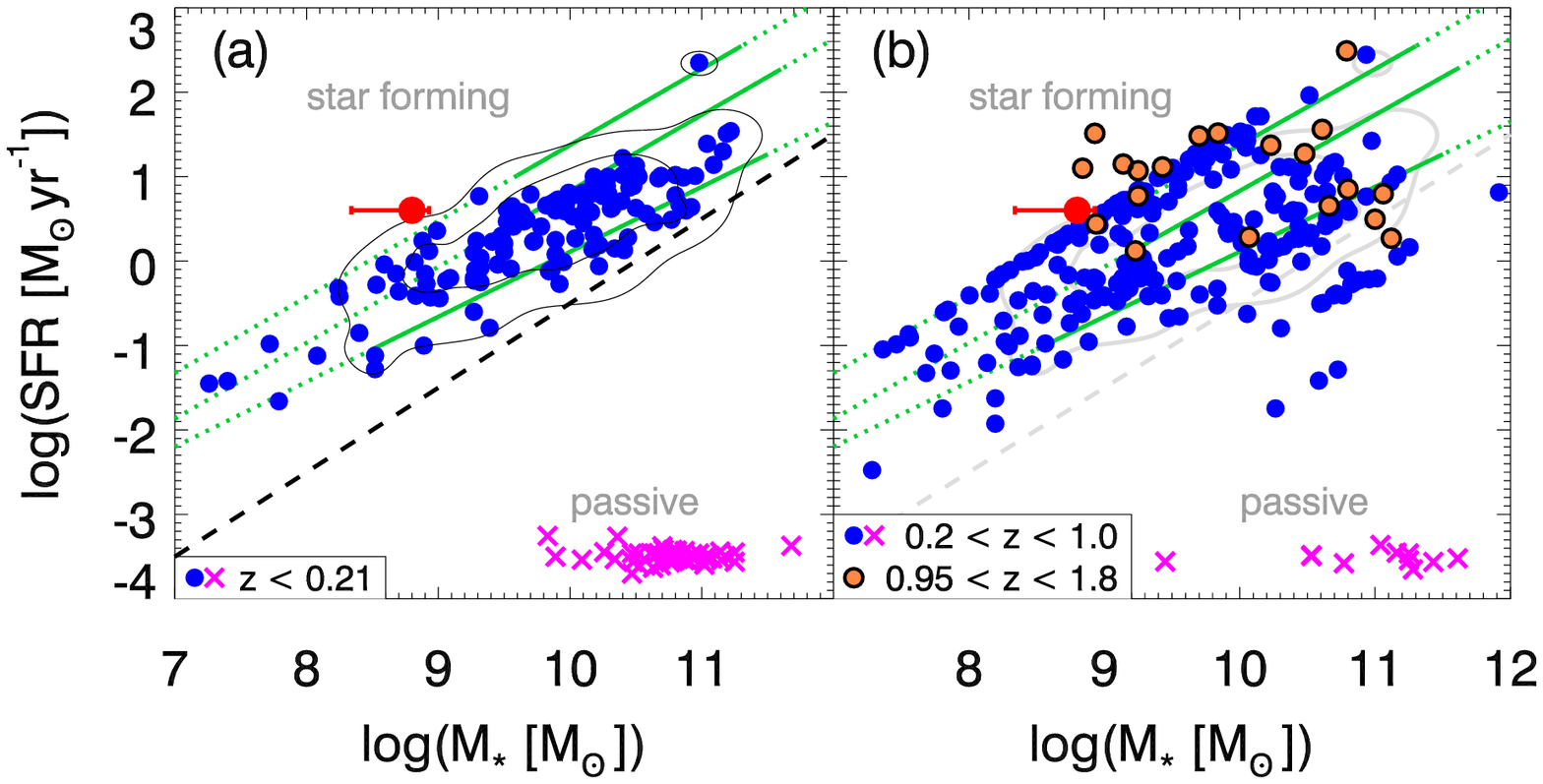}
	\caption{
	\red{The SFR-mass relation for SN Ia host galaxies.
	The (red) asterisk denotes the host of SN Primo with error bars.
	Filled (blue) circles mark star-forming hosts in each sample.
	The (magenta) crosses marks the passive hosts in each sample.}
	The solid and dotted (green) lines show the correlation between SFR and stellar mass for $z\sim0$ (bottom), $z=1$ (middle), and $z=2$ (top) from \citet[$z=2$]{2007ApJ...670..156D} and \citet[$z\sim0$ and $z=1$]{2007A&A...468...33E}. The solid section of each line marks the range of validity of the relations.
	(a) The low redshift sample ($z<0.21$) from SDSS \citep{2010ApJ...722..566L}.
	The dashed line marks the cut, $\log(sSFR)=-10.6$, between star forming and passive galaxies.
	The contours mark the region enclosing 68\% and 95\% of the star-forming sample.
	\red{(b) The high redshift samples from HST \citep[$0.95<z<1.8$, open (orange) circles]{2011ApJ...731...72T} and SNLS \citep[$0.2 < z<1.0$, filled (blue) circles / (magenta) crosses]{2010MNRAS.406..782S}. The apparent upper diagonal ridge-line for the \citet{2010MNRAS.406..782S} data is due to shortcomings in their SED fitting.
	}
	\label{fig_sfr_mass}}
\end{figure}
\begin{figure}
	\epsscale{0.75}
	\plotone{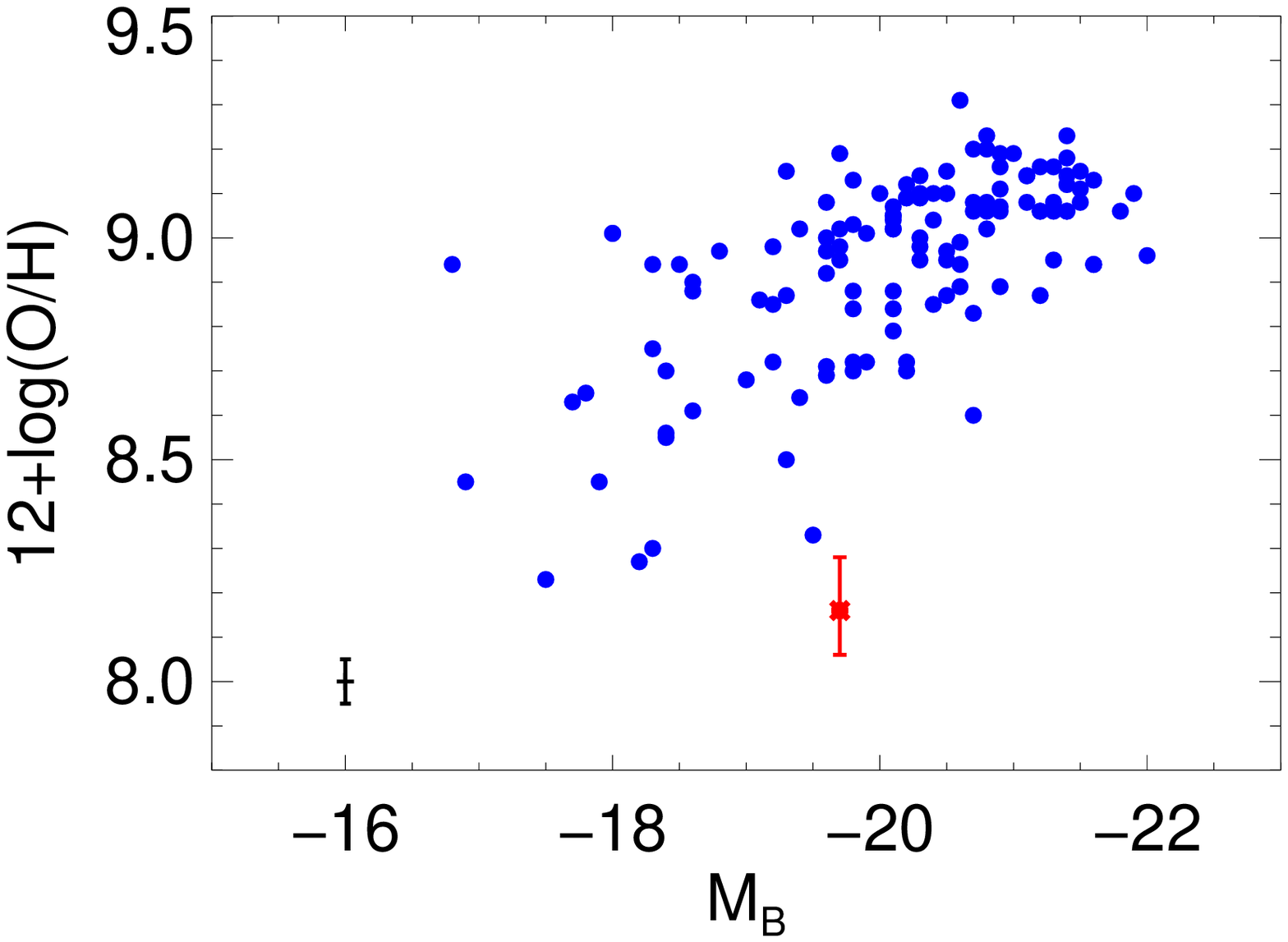}
	\caption{
	Metallicity--luminosity relation for SN Ia host galaxies. The (red) asterisk denotes the host of SN Primo with error bars. The (blue) filled circles denotes the sample of \citet[$z<0.04$, \red{median error bar plotter to the lower left}]{2008ApJ...673..999P}. 
	\label{fig_LZ}}
\end{figure}
\red{In Figure \ref{fig_sfr_mass} we plot the SFR vs.\ stellar mass for the host of SN Primo in comparison to both a low redshift \citep[$z<0.21$]{2010ApJ...722..566L} and two high redshift samples from HST \citep[$0.95<z<1.8$]{2011ApJ...731...72T} and SNLS \citep[$0.2 < z<1.0$]{2010MNRAS.406..782S} samples.}
The host of SN Primo clearly stands out from the \red{low-$z$} sample, due to its high specific star-formation rate. The relation between SFR and stellar mass is expected to evolve with redshift as seen in observations \citep{2007ApJ...670..156D,2007A&A...468...33E}. If SN host galaxies are representative of field galaxies the blue points in Figure \ref{fig_sfr_mass} are  expected to shift upwards in the same way as the green dashed lines (signifying $z=0$, 1, and 2).
\red{It is hard to see that trend from the SNLS and HST samples due to the scatter (HST) and degeneracies (SNLS) in the data. The degeneracies (at constant sSFR) in the SNLS data is due to shortcomings in the SED fitting in \citet{2010MNRAS.406..782S}.}

Its derived metallicity is not unusually low for galaxies in general, but the metallicity is very low for SN Ia host galaxies \citep{2005ApJ...634..210G,2008ApJ...673..999P}. We plot the metallicity-luminosity relation for the sample of \citet[$z<0.04$]{2008ApJ...673..999P}.
The host of SN Primo has a lower metallicity than any of the \red{low-$z$} galaxies (see Figure \ref{fig_LZ}).
The host of SN Primo also falls below the mass-metallicity relation \citep{2004ApJ...613..898T}. We check why this could be the case by comparing the host of SN Primo to the Fundamental Metallicity Relation (FMR) of star-forming galaxies \citep{2010A&A...521L..53L,2010MNRAS.408.2115M} which relates metallicity, stellar mass, and SFR. \citet{2011MNRAS.414.1263M} updated the low-mass slope of the FMR relation using GRB host galaxies. The metallicity predicted by the FMR relation is within the error bars of the measured metallicity. The residual between SN Primo and the FMR relation is $\Delta[12 + \log(\frac{O}{H})] = 0.07\pm0.15$. The host of SN Primo is therefore consistent with the FMR relation defined in \citet{2011MNRAS.414.1263M}. The FMR relation is consistent with a simple model \citep{2012arXiv1202.4770D} where the balance of gas infall, outflow, and star formation brings out the relation between SFR, metallicity and stellar mass seen in the FMR relation.

The stellar age of $\sim10^{8.6}$ years (Table \ref{tab_sed}) could give an upper limit on the delay time of SN Primo, assuming there is no underlying old stellar population.
This would put SN Primo in the prompt progenitor distribution \citep[see][for a review]{2006ApJ...648..868S}.
There are however caveats to the values derived from our SED fitting. It is assumed that there is no underlying old stellar population, which can not be ruled out. This is also seen in Table \ref{tab_sed} where ages up to $\sim10^{9}$ are still consistent within 1$\sigma$.

\red{In this paper we have used a redshift independent definition of a starburst based on the value of the sSFR \citep[$\log(sSFR)>-9.5$,][]{2006ApJ...648..868S}. Alternatively a starburst can be defined based on the SFR and $M_*$ of galaxies at the same redshift --- the so called main-sequence (MS) of galaxies (indicated in Figure \ref{fig_sfr_mass}a).
The evolution of the MS with redshift, however, is not fully settled \citep[see][among other]{2007ApJ...670..156D,2011ApJ...742...96W,2012ApJ...754L..29W}. As indicated in Figure \ref{fig_sfr_mass}a  the MS fits at $z=1$ and $z=2$ would have to be extrapolated down to the mass of the host of SN Primo.}

\citet{2005ApJ...634..210G} showed that the light-curve shape correlates with the Hubble type of the host galaxy and \citet{2012ApJ...750....1M} showed that both light-curve shape and SN peak color are different between early-type and late-type galaxies.
\citet{2011ApJ...737..102S} showed by splitting up the SNLS3 sample of SNe Ia into a high and low sSFR sample, that the host galaxy has an influence on the mean SN peak brightness and the correction of light-curve shape and color correction. Galaxy evolution models find that sSFR increases with redshift out to at least $z=2$.
Using the mass of the host of SN Primo the ``host term'' of \citet{2010ApJ...715..743K} is 0.3 mag (super-luminous SN).
These dependencies highlight that the bulk of the training sample of SNe Ia lies below $z<1$ where the host galaxies are older and in general have a smaller sSFR. As a consequence, this could introduce a potential bias in the distances derived to the high sSFR host galaxies, when not explicitly including the host correction. As the sample of high redshift and high sSFR SNe grows the size of this effect can be investigated further.

\acknowledgments
We thank Robert Kirshner, Mark Dickinson, Peter M. Garnavich, Brian Hayden, Giorgos Leloudas for comments and discussions.
We thank the anonymous referee for valuable comments that helped improve this manuscript.
We thank Martin Sparre for providing his X-shooter meta-pipeline, which has simplified the reduction of the X-shooter spectra significantly.
This work is based on observations taken by the CANDELS Multi-Cycle Treasury Program with the NASA/ESA HST, which is operated by the Association of Universities for Research in Astronomy, Inc., under NASA contract NAS5-26555.
The Dark Cosmology Centre is funded by the Danish National Research Foundation.
This research has made use of the NASA/IPAC Extragalactic Database (NED) which is operated by the Jet Propulsion Laboratory, California Institute of Technology, under contract with the National Aeronautics and Space Administration.



{\it Facilities:} \facility{VLT:Kueyen (X-shooter)}, \facility{HST (ACS, WCF3)}, \facility{VLT:Melipal (ISAAC, VIMOS)}, \facility{Spitzer (IRAC)}


\begin{deluxetable}{llrrc}
	\tablecaption{Photometry of the host of SN Primo\label{tab_photometry}}
	\tablehead{
		\colhead{Filter} &
		\colhead{Instrument} &
		\colhead{$\lambda_{\rm eff}$} &
		\colhead{Magnitude\tablenotemark{a}} &
		\colhead{Corrected\tablenotemark{b}} \\
		\colhead{} &
		\colhead{} &
		\colhead{(nm)} &
		\colhead{(AB mag)} &
		\colhead{(AB mag)}
	}
	\startdata
	U-band	&	VLT/VIMOS	&	375.3 & $24.69\pm0.02$ & \ldots \\
	F435W (B) &	{\it HST}/ACS &	432.8 & $24.84\pm0.02$ & \ldots \\
	F606W (V) &	{\it HST}/ACS &	595.8 & $24.84\pm0.02$ & \ldots \\
	F775W (i) &	{\it HST}/ACS &	770.6 & $24.86\pm0.03$ & \ldots \\
	F850LP (z) &	{\it HST}/ACS &	905.3 & $24.80\pm0.04$ & $24.88\pm0.04$ \\
	F105W (Y) &	{\it HST}/WFC3 &	1059 & $24.50\pm0.01$ & $24.56\pm0.01$ \\
	F125W (J) &	{\it HST}/WFC3 &	1252 & $24.27\pm0.01$ & $24.71\pm0.01$ \\
	F160W (H) &	{\it HST}/WFC3 &	1544 & $24.36\pm0.01$ & $24.71\pm0.01$ \\
	$\rm K_s$-band	&	VLT/ISAAC	&	2168 & $24.47\pm0.17$ & \ldots \\
	Channel 1 &	{\it Spitzer}/IRAC &	3563 & $24.43\pm0.05$ & \ldots \\
	Channel 2 &	{\it Spitzer}/IRAC &	4511 & $24.49\pm0.09$ & \ldots \\
	\enddata
	\footnotesize
	\tablenotetext{a}{The magnitudes before subtraction of the emission-line fluxes.}
	\tablenotetext{b}{The magnitudes after subtraction of the emission-line fluxes.
	}
\end{deluxetable}

\begin{deluxetable}{lrrrl}
	\tablecaption{Summary of SED fitting using FAST.\label{tab_sed}}
	\tablehead{
		\colhead{Parameter\tablenotemark{a}} &
		\colhead{Exponential\tablenotemark{b}} &
		\colhead{Delayed\tablenotemark{c}} &
		\colhead{Truncated\tablenotemark{d}}
	}
	\startdata
\smallskip
$\log(t_* {\rm [yr]})$ & $ 8.50^{+0.28}_{-1.41}$ & $ 8.60^{+0.51}_{-1.25}$ & $ 8.65^{+0.13}_{-1.20}$ \\
\smallskip
$\log(\tau {\rm [yr]})$ & $ 8.80^{+2.20}_{-2.30}$ & $ 8.30^{+2.70}_{-1.62}$ & $ 9.40^{+1.60}_{-1.96}$ \\
\smallskip
$Z$ & $ 0.020^{+0.023}_{-0.016}$ & $ 0.020^{+0.021}_{-0.016}$ & $ 0.020^{+0.015}_{-0.016}$ \\
\smallskip
$A_V$ & $ 0.00^{+0.71}_{-0.00}$ & $ 0.00^{+0.71}_{-0.00}$ & $ 0.00^{+0.70}_{-0.00}$ \\
\tableline
\smallskip
$\log(M_* {\rm [M_\odot]})$ & $ 8.80^{+0.13}_{-0.46}$ & $ 8.81^{+0.16}_{-0.34}$ & $ 8.84^{+0.07}_{-0.39}$ \\
\smallskip
$\log(SFR {\rm [M_\odot yr^{-1}]})$ & $ 0.33^{+0.77}_{-0.87}$ & $ 0.32^{+0.84}_{-0.49}$ & $ 0.35^{+0.73}_{-99.35}$ \\
\smallskip
$\log(sSFR {\rm [yr^{-1}]})$ & $ -8.47^{+1.07}_{-0.61}$ & $ -8.49^{+1.15}_{-0.27}$ & $ -8.50^{+1.06}_{-90.50}$ \\
\smallskip
$\log(t_* / \tau)$ & $ -0.30^{+1.05}_{-3.25}$ & $ 0.30^{+0.50}_{-3.60}$ & $ -0.75^{+0.80}_{-2.75}$ \\
\tableline
$\tilde\chi^2$ &  $19.9$ &  $19.8$ &  $19.8$ \\
	\enddata
	\footnotesize
	\tablenotetext{a}{We have assumed a Chabrier IMF}
	\tablenotetext{b}{Exponential star-formation history: $SFR(t)\propto \exp(-t/\tau)$, $\tau>0$}
	\tablenotetext{c}{Delayed star-formation history: $SFR(t)\propto t \cdot \exp(-t/\tau)$}
	\tablenotetext{d}{Truncated star-formation history: $SFR(t) =$ constant; for $t \in [t_*, t_*+\tau]$, else $0$}
\end{deluxetable}
\begin{deluxetable}{lllll}
	\tablecaption{Emission lines detected in the spectrum of the host of SN Primo.\label{tab_lines}}
	\tablehead{
		\colhead{Line} &
		\colhead{Wavelength} &
		\multicolumn{2}{c}{Observed FWHM} &
		\colhead{Flux\tablenotemark{a}} \\ 
		&
		\colhead{[nm]} &
		\colhead{[nm]} &
		\colhead{[km s$^{-1}$]} &
		\colhead{[$10^{-17}$ erg s$^{-1}$ cm$^{-2}$]}
	}
	\startdata
		H$\alpha$ $\lambda$6563 &
		$1673.43\phd^{+0.01}_{-0.02}$ &
		$0.65\phd^{+0.05}_{-0.04}$ &
		$117\phd^{ +8}_{ -7}$ &
		\phm{< }\phn$ 5.1\phd^{ +0.3}_{ -0.3}$ \\

		[\ion{O}{3}] $\lambda$5007 &
		$1276.71\phd^{+0.02}_{-0.02}$ &
		$0.44\phd^{+0.07}_{-0.05}$ &
		$103\phd^{+16}_{-11}$ &
		\phm{< }\phn$ 6.0\phd^{ +0.6}_{ -0.5}$ \\

		[\ion{O}{3}] $\lambda$4959 &
		$1264.43\phd^{+0.04}_{-0.03}$ &
		$0.31\phd^{+0.07}_{-0.08}$ &
		\phn$ 75\phd^{+16}_{-17}$ &
		\phm{< }\phn$ 2.1\phd^{ +0.4}_{ -0.4}$ \\

		H$\beta$ $\lambda$4861\tablenotemark{b} &
		\multicolumn{1}{c}{-} &
		\multicolumn{1}{c}{-} &
		\multicolumn{1}{c}{-} &
		\phm{< }\phn$ 1.4\phd^{ +0.4}_{ -0.5}$ \\

		[\ion{O}{2}] $\lambda\lambda$3726,3729\tablenotemark{c} &
		\phn$ 950.90\phd^{+0.08}_{-0.05}$ &
		$0.35\phd^{+0.27}_{-0.07}$ &
		$112\phd^{+86}_{-21}$ &
		\phm{< }\phn$ 2.6\phd^{ +1.0}_{ -0.8}$ \\
		\tableline
		[\ion{N}{2}] $\lambda$6586\tablenotemark{d} &
		\multicolumn{1}{c}{-} &
		\multicolumn{1}{c}{-} &
		\multicolumn{1}{c}{-} &
		\red{$<0.3$} \\

		Ly$\alpha$ $\lambda$1216\tablenotemark{e} &
		\multicolumn{1}{c}{-} &
		\multicolumn{1}{c}{-} &
		\multicolumn{1}{c}{-} &
		\phm{< }$11.2\phd^{ +4.0}_{ -3.9}$ \\
	\enddata
	\footnotesize
	\tablenotetext{a}{Fit of the observed flux corrected for Galactic extinction ($E(B-V)=0.008$).}
	\tablenotetext{b}{In each resampling the following fit was performed: The central-wavelength of the line was fixed to $\lambda_{\rm H\beta}(1+z)$, where $\lambda_{\rm H\beta}=486.1325 \rm{nm}$ and $z$ is the redshift. FWHM(H$\beta$) was fixed to FWHM(H$\alpha$) in velocity units. Only the peak intensity was allowed to vary.}
	\tablenotetext{c}{The wavelength is that of [\ion{O}{2}] ${\lambda3729}$ (the red component) only, the flux is the sum of both components.}
	\tablenotetext{d}{$5\sigma$ upper limit of the non-detection.}
	\tablenotetext{e}{The Ly$\alpha$ flux is the co-added flux from $v=0-600$ km s$^{-1}$. The error bars only cover the statistical errors. The systematic error is $\sim40\%$.}
\end{deluxetable}
\begin{deluxetable}{lrl}
	\tablecaption{Spectroscopic summary.\label{tab_results}}
	\tablehead{
		\colhead{Parameter} &
		\colhead{Value} &
		\colhead{Assumed IMF}
	}
	\startdata
		\smallskip
		Redshift (heliocentric)& $z$ = $1.54992^{+0.00008}_{-0.00004}$\\
		\smallskip
		Metallicity & $12+\log(\frac{O}{H}) = 8.12^{+0.09}_{-0.10}$ \\
		\smallskip
		& $Z=0.27^{+0.06}_{-0.06} Z_\odot$\tablenotemark{a}\\
		Extinction & $A_V =0.6^{+1.1}_{-0.7}$ mag\\
		Ly$\alpha$ escape fraction & $f_{esc}=0.25\pm0.09$ $(\pm0.10)$\tablenotemark{b} \\
		\tableline
		\multicolumn{2}{l}{Star-formation rate:}\\
		& $SFR = 6.4\pm0.3$ M$_\odot$ yr$^{-1}$ & Salpeter \\
		& $SFR = 4.3\pm0.2$ M$_\odot$ yr$^{-1}$ & Kroupa \\
		& $SFR = 3.8\pm0.2$ M$_\odot$ yr$^{-1}$ & Chabrier \\
		\tableline
		\multicolumn{2}{l}{Specific star-formation rate:}\\
		& $\log(sSFR[{\rm yr}^{-1}]) = -7.8 \pm 0.2$ & Salpeter \\
		& $\log(sSFR[{\rm yr}^{-1}]) = -8.0 \pm 0.2$ & Kroupa \\
		& $\log(sSFR[{\rm yr}^{-1}]) = -8.1 \pm 0.2$ & Chabrier \\
	\enddata
	\footnotesize
	\tablenotetext{a}{Assuming a solar oxygen abundance of 8.69 \citep{2009ARA&A..47..481A}}
	\tablenotetext{b}{The value in parenthesis covers the systematic uncertainty on the flux of Ly$\alpha$ of 40\%.}
\end{deluxetable}

\appendix





\bibliography{/Users/teddy/Documents/bibtexDB.bib}{}

\begin{thebibliography}{61}
\expandafter\ifx\csname natexlab\endcsname\relax\def\natexlab#1{#1}\fi

\bibitem[{{Asplund} {et~al.}(2009){Asplund}, {Grevesse}, {Sauval}, \&
  {Scott}}]{2009ARA&A..47..481A}
{Asplund}, M., {Grevesse}, N., {Sauval}, A.~J., \& {Scott}, P. 2009, \araa, 47,
  481

\bibitem[{{Atek} {et~al.}(2009){Atek}, {Kunth}, {Schaerer}, {Hayes},
  {Deharveng}, {{\"O}stlin}, \& {Mas-Hesse}}]{2009A&A...506L...1A}
{Atek}, H., {Kunth}, D., {Schaerer}, D., {Hayes}, M., {Deharveng}, J.~M.,
  {{\"O}stlin}, G., \& {Mas-Hesse}, J.~M. 2009, \aap, 506, L1

\bibitem[{{Baldwin} {et~al.}(1981){Baldwin}, {Phillips}, \&
  {Terlevich}}]{1981PASP...93....5B}
{Baldwin}, J.~A., {Phillips}, M.~M., \& {Terlevich}, R. 1981, \pasp, 93, 5

\bibitem[{{Brinchmann} {et~al.}(2004){Brinchmann}, {Charlot}, {White},
  {Tremonti}, {Kauffmann}, {Heckman}, \& {Brinkmann}}]{2004MNRAS.351.1151B}
{Brinchmann}, J., {Charlot}, S., {White}, S.~D.~M., {Tremonti}, C.,
  {Kauffmann}, G., {Heckman}, T., \& {Brinkmann}, J. 2004, \mnras, 351, 1151

\bibitem[{{Brocklehurst}(1971)}]{1971MNRAS.153..471B}
{Brocklehurst}, M. 1971, \mnras, 153, 471

\bibitem[{{Bruzual} \& {Charlot}(2003)}]{2003MNRAS.344.1000B}
{Bruzual}, G., \& {Charlot}, S. 2003, \mnras, 344, 1000

\bibitem[{{Calzetti}(2001)}]{2001PASP..113.1449C}
{Calzetti}, D. 2001, \pasp, 113, 1449

\bibitem[{{Cardelli} {et~al.}(1989){Cardelli}, {Clayton}, \&
  {Mathis}}]{1989ApJ...345..245C}
{Cardelli}, J.~A., {Clayton}, G.~C., \& {Mathis}, J.~S. 1989, \apj, 345, 245

\bibitem[{{Chabrier}(2003)}]{2003PASP..115..763C}
{Chabrier}, G. 2003, \pasp, 115, 763

\bibitem[{{Conley} {et~al.}(2011){Conley}, {Guy}, {Sullivan}, {Regnault},
  {Astier}, {Balland}, {Basa}, {Carlberg}, {Fouchez}, {Hardin}, {Hook},
  {Howell}, {Pain}, {Palanque-Delabrouille}, {Perrett}, {Pritchet}, {Rich},
  {Ruhlmann-Kleider}, {Balam}, {Baumont}, {Ellis}, {Fabbro}, {Fakhouri},
  {Fourmanoit}, {Gonz{\'a}lez-Gait{\'a}n}, {Graham}, {Hudson}, {Hsiao},
  {Kronborg}, {Lidman}, {Mourao}, {Neill}, {Perlmutter}, {Ripoche}, {Suzuki},
  \& {Walker}}]{2011ApJS..192....1C}
{Conley}, A., {et~al.} 2011, \apjs, 192, 1

\bibitem[{{Daddi} {et~al.}(2007){Daddi}, {Dickinson}, {Morrison}, {Chary},
  {Cimatti}, {Elbaz}, {Frayer}, {Renzini}, {Pope}, {Alexander}, {Bauer},
  {Giavalisco}, {Huynh}, {Kurk}, \& {Mignoli}}]{2007ApJ...670..156D}
{Daddi}, E., {et~al.} 2007, \apj, 670, 156

\bibitem[{{Dahlen} {et~al.}(2010){Dahlen}, {Mobasher}, {Dickinson}, {Ferguson},
  {Giavalisco}, {Grogin}, {Guo}, {Koekemoer}, {Lee}, {Lee}, {Nonino}, {Riess},
  \& {Salimbeni}}]{2010ApJ...724..425D}
{Dahlen}, T., {et~al.} 2010, \apj, 724, 425

\bibitem[{{Dayal} {et~al.}(2012){Dayal}, {Ferrara}, \&
  {Dunlop}}]{2012arXiv1202.4770D}
{Dayal}, P., {Ferrara}, A., \& {Dunlop}, J.~S. 2012, ArXiv e-prints,
  arXiv:1202.4770

\bibitem[{{D'Odorico} {et~al.}(2006){D'Odorico}, {Dekker}, {Mazzoleni},
  {Vernet}, {Guinouard}, {Groot}, {Hammer}, {Rasmussen}, {Kaper}, {Navarro},
  {Pallavicini}, {Peroux}, \& {Zerbi}}]{2006SPIE.6269E..98D}
{D'Odorico}, S., {et~al.} 2006, in Society of Photo-Optical Instrumentation
  Engineers (SPIE) Conference Series, Vol. 6269

\bibitem[{{Elbaz} {et~al.}(2007){Elbaz}, {Daddi}, {Le Borgne}, {Dickinson},
  {Alexander}, {Chary}, {Starck}, {Brandt}, {Kitzbichler}, {MacDonald},
  {Nonino}, {Popesso}, {Stern}, \& {Vanzella}}]{2007A&A...468...33E}
{Elbaz}, D., {et~al.} 2007, \aap, 468, 33

\bibitem[{{Gallagher} {et~al.}(2005){Gallagher}, {Garnavich}, {Berlind},
  {Challis}, {Jha}, \& {Kirshner}}]{2005ApJ...634..210G}
{Gallagher}, J.~S., {Garnavich}, P.~M., {Berlind}, P., {Challis}, P., {Jha},
  S., \& {Kirshner}, R.~P. 2005, \apj, 634, 210

\bibitem[{{Grogin} {et~al.}(2011){Grogin}, {Kocevski}, {Faber}, {Ferguson},
  {Koekemoer}, {Riess}, {Acquaviva}, {Alexander}, {Almaini}, {Ashby}, {Barden},
  {Bell}, {Bournaud}, {Brown}, {Caputi}, {Casertano}, {Cassata}, {Castellano},
  {Challis}, {Chary}, {Cheung}, {Cirasuolo}, {Conselice}, {Roshan Cooray},
  {Croton}, {Daddi}, {Dahlen}, {Dav{\'e}}, {de Mello}, {Dekel}, {Dickinson},
  {Dolch}, {Donley}, {Dunlop}, {Dutton}, {Elbaz}, {Fazio}, {Filippenko},
  {Finkelstein}, {Fontana}, {Gardner}, {Garnavich}, {Gawiser}, {Giavalisco},
  {Grazian}, {Guo}, {Hathi}, {H{\"a}ussler}, {Hopkins}, {Huang}, {Huang},
  {Jha}, {Kartaltepe}, {Kirshner}, {Koo}, {Lai}, {Lee}, {Li}, {Lotz}, {Lucas},
  {Madau}, {McCarthy}, {McGrath}, {McIntosh}, {McLure}, {Mobasher},
  {Moustakas}, {Mozena}, {Nandra}, {Newman}, {Niemi}, {Noeske}, {Papovich},
  {Pentericci}, {Pope}, {Primack}, {Rajan}, {Ravindranath}, {Reddy}, {Renzini},
  {Rix}, {Robaina}, {Rodney}, {Rosario}, {Rosati}, {Salimbeni}, {Scarlata},
  {Siana}, {Simard}, {Smidt}, {Somerville}, {Spinrad}, {Straughn}, {Strolger},
  {Telford}, {Teplitz}, {Trump}, {van der Wel}, {Villforth}, {Wechsler},
  {Weiner}, {Wiklind}, {Wild}, {Wilson}, {Wuyts}, {Yan}, \&
  {Yun}}]{2011ApJS..197...35G}
{Grogin}, N.~A., {et~al.} 2011, \apjs, 197, 35

\bibitem[{{Hamuy} {et~al.}(2000){Hamuy}, {Trager}, {Pinto}, {Phillips},
  {Schommer}, {Ivanov}, \& {Suntzeff}}]{2000AJ....120.1479H}
{Hamuy}, M., {Trager}, S.~C., {Pinto}, P.~A., {Phillips}, M.~M., {Schommer},
  R.~A., {Ivanov}, V., \& {Suntzeff}, N.~B. 2000, \aj, 120, 1479

\bibitem[{{Hayes} {et~al.}(2011){Hayes}, {Schaerer}, {{\"O}stlin}, {Mas-Hesse},
  {Atek}, \& {Kunth}}]{2011ApJ...730....8H}
{Hayes}, M., {Schaerer}, D., {{\"O}stlin}, G., {Mas-Hesse}, J.~M., {Atek}, H.,
  \& {Kunth}, D. 2011, \apj, 730, 8

\bibitem[{{Hillebrandt} \& {Niemeyer}(2000)}]{2000ARA&A..38..191H}
{Hillebrandt}, W., \& {Niemeyer}, J.~C. 2000, \araa, 38, 191

\bibitem[{{Kelly} {et~al.}(2010){Kelly}, {Hicken}, {Burke}, {Mandel}, \&
  {Kirshner}}]{2010ApJ...715..743K}
{Kelly}, P.~L., {Hicken}, M., {Burke}, D.~L., {Mandel}, K.~S., \& {Kirshner},
  R.~P. 2010, \apj, 715, 743

\bibitem[{{Kennicutt}(1998)}]{1998ARA&A..36..189K}
{Kennicutt}, Jr., R.~C. 1998, \araa, 36, 189

\bibitem[{{Kewley} \& {Dopita}(2002)}]{2002ApJS..142...35K}
{Kewley}, L.~J., \& {Dopita}, M.~A. 2002, \apjs, 142, 35

\bibitem[{{Kewley} \& {Ellison}(2008)}]{2008ApJ...681.1183K}
{Kewley}, L.~J., \& {Ellison}, S.~L. 2008, \apj, 681, 1183

\bibitem[{{Kirshner}(2010)}]{2010deot.book..151K}
{Kirshner}, R.~P. 2010, {Foundations of supernova cosmology}, ed.
  P.~{Ruiz-Lapuente}, 151

\bibitem[{{Kobulnicky} \& {Kewley}(2004)}]{2004ApJ...617..240K}
{Kobulnicky}, H.~A., \& {Kewley}, L.~J. 2004, \apj, 617, 240

\bibitem[{{Koekemoer} {et~al.}(2011){Koekemoer}, {Faber}, {Ferguson}, {Grogin},
  {Kocevski}, {Koo}, {Lai}, {Lotz}, {Lucas}, {McGrath}, {Ogaz}, {Rajan},
  {Riess}, {Rodney}, {Strolger}, {Casertano}, {Castellano}, {Dahlen},
  {Dickinson}, {Dolch}, {Fontana}, {Giavalisco}, {Grazian}, {Guo}, {Hathi},
  {Huang}, {van der Wel}, {Yan}, {Acquaviva}, {Alexander}, {Almaini}, {Ashby},
  {Barden}, {Bell}, {Bournaud}, {Brown}, {Caputi}, {Cassata}, {Challis},
  {Chary}, {Cheung}, {Cirasuolo}, {Conselice}, {Roshan Cooray}, {Croton},
  {Daddi}, {Dav{\'e}}, {de Mello}, {de Ravel}, {Dekel}, {Donley}, {Dunlop},
  {Dutton}, {Elbaz}, {Fazio}, {Filippenko}, {Finkelstein}, {Frazer}, {Gardner},
  {Garnavich}, {Gawiser}, {Gruetzbauch}, {Hartley}, {H{\"a}ussler},
  {Herrington}, {Hopkins}, {Huang}, {Jha}, {Johnson}, {Kartaltepe},
  {Khostovan}, {Kirshner}, {Lani}, {Lee}, {Li}, {Madau}, {McCarthy},
  {McIntosh}, {McLure}, {McPartland}, {Mobasher}, {Moreira}, {Mortlock},
  {Moustakas}, {Mozena}, {Nandra}, {Newman}, {Nielsen}, {Niemi}, {Noeske},
  {Papovich}, {Pentericci}, {Pope}, {Primack}, {Ravindranath}, {Reddy},
  {Renzini}, {Rix}, {Robaina}, {Rosario}, {Rosati}, {Salimbeni}, {Scarlata},
  {Siana}, {Simard}, {Smidt}, {Snyder}, {Somerville}, {Spinrad}, {Straughn},
  {Telford}, {Teplitz}, {Trump}, {Vargas}, {Villforth}, {Wagner}, {Wandro},
  {Wechsler}, {Weiner}, {Wiklind}, {Wild}, {Wilson}, {Wuyts}, \&
  {Yun}}]{2011ApJS..197...36K}
{Koekemoer}, A.~M., {et~al.} 2011, \apjs, 197, 36

\bibitem[{{Komatsu} {et~al.}(2011){Komatsu}, {Smith}, {Dunkley}, {Bennett},
  {Gold}, {Hinshaw}, {Jarosik}, {Larson}, {Nolta}, {Page}, {Spergel},
  {Halpern}, {Hill}, {Kogut}, {Limon}, {Meyer}, {Odegard}, {Tucker}, {Weiland},
  {Wollack}, \& {Wright}}]{2011ApJS..192...18K}
{Komatsu}, E., {et~al.} 2011, \apjs, 192, 18

\bibitem[{{Kriek} {et~al.}(2009){Kriek}, {van Dokkum}, {Labb{\'e}}, {Franx},
  {Illingworth}, {Marchesini}, \& {Quadri}}]{2009ApJ...700..221K}
{Kriek}, M., {van Dokkum}, P.~G., {Labb{\'e}}, I., {Franx}, M., {Illingworth},
  G.~D., {Marchesini}, D., \& {Quadri}, R.~F. 2009, \apj, 700, 221

\bibitem[{{Laidler} {et~al.}(2007){Laidler}, {Papovich}, {Grogin}, {Idzi},
  {Dickinson}, {Ferguson}, {Hilbert}, {Clubb}, \&
  {Ravindranath}}]{2007PASP..119.1325L}
{Laidler}, V.~G., {et~al.} 2007, \pasp, 119, 1325

\bibitem[{{Lampeitl} {et~al.}(2010){Lampeitl}, {Smith}, {Nichol}, {Bassett},
  {Cinabro}, {Dilday}, {Foley}, {Frieman}, {Garnavich}, {Goobar}, {Im}, {Jha},
  {Marriner}, {Miquel}, {Nordin}, {{\"O}stman}, {Riess}, {Sako}, {Schneider},
  {Sollerman}, \& {Stritzinger}}]{2010ApJ...722..566L}
{Lampeitl}, H., {et~al.} 2010, \apj, 722, 566

\bibitem[{{Lara-L{\'o}pez} {et~al.}(2010){Lara-L{\'o}pez}, {Cepa},
  {Bongiovanni}, {P{\'e}rez Garc{\'{\i}}a}, {Ederoclite}, {Casta{\~n}eda},
  {Fern{\'a}ndez Lorenzo}, {Povi{\'c}}, \&
  {S{\'a}nchez-Portal}}]{2010A&A...521L..53L}
{Lara-L{\'o}pez}, M.~A., {et~al.} 2010, \aap, 521, L53+

\bibitem[{{Mannucci} {et~al.}(2010){Mannucci}, {Cresci}, {Maiolino}, {Marconi},
  \& {Gnerucci}}]{2010MNRAS.408.2115M}
{Mannucci}, F., {Cresci}, G., {Maiolino}, R., {Marconi}, A., \& {Gnerucci}, A.
  2010, \mnras, 408, 2115

\bibitem[{{Mannucci} {et~al.}(2011){Mannucci}, {Salvaterra}, \&
  {Campisi}}]{2011MNRAS.414.1263M}
{Mannucci}, F., {Salvaterra}, R., \& {Campisi}, M.~A. 2011, \mnras, 414, 1263

\bibitem[{{March} {et~al.}(2011){March}, {Trotta}, {Berkes}, {Starkman}, \&
  {Vaudrevange}}]{2011MNRAS.418.2308M}
{March}, M.~C., {Trotta}, R., {Berkes}, P., {Starkman}, G.~D., \&
  {Vaudrevange}, P.~M. 2011, \mnras, 418, 2308

\bibitem[{{McGaugh}(1991)}]{1991ApJ...380..140M}
{McGaugh}, S.~S. 1991, \apj, 380, 140

\bibitem[{{Meyers} {et~al.}(2012){Meyers}, {Aldering}, {Barbary}, {Barrientos},
  {Brodwin}, {Dawson}, {Deustua}, {Doi}, {Eisenhardt}, {Faccioli}, {Fakhouri},
  {Fruchter}, {Gilbank}, {Gladders}, {Goldhaber}, {Gonzalez}, {Hattori},
  {Hsiao}, {Ihara}, {Kashikawa}, {Koester}, {Konishi}, {Lidman}, {Lubin},
  {Morokuma}, {Oda}, {Perlmutter}, {Postman}, {Ripoche}, {Rosati}, {Rubin},
  {Rykoff}, {Spadafora}, {Stanford}, {Suzuki}, {Takanashi}, {Tokita}, {Yasuda},
  \& {Supernova Cosmology Project}}]{2012ApJ...750....1M}
{Meyers}, J., {et~al.} 2012, \apj, 750, 1

\bibitem[{{Oesch} {et~al.}(2010){Oesch}, {Bouwens}, {Illingworth}, {Carollo},
  {Franx}, {Labb{\'e}}, {Magee}, {Stiavelli}, {Trenti}, \& {van
  Dokkum}}]{2010ApJ...709L..16O}
{Oesch}, P.~A., {et~al.} 2010, \apjl, 709, L16

\bibitem[{{Osterbrock} \& {Ferland}(2006)}]{2006agna.book.....O}
{Osterbrock}, D.~E., \& {Ferland}, G.~J. 2006, {Astrophysics of gaseous nebulae
  and active galactic nuclei} (University Science Books)

\bibitem[{{Perlmutter} {et~al.}(1999){Perlmutter}, {Aldering}, {Goldhaber},
  {Knop}, {Nugent}, {Castro}, {Deustua}, {Fabbro}, {Goobar}, {Groom}, {Hook},
  {Kim}, {Kim}, {Lee}, {Nunes}, {Pain}, {Pennypacker}, {Quimby}, {Lidman},
  {Ellis}, {Irwin}, {McMahon}, {Ruiz-Lapuente}, {Walton}, {Schaefer}, {Boyle},
  {Filippenko}, {Matheson}, {Fruchter}, {Panagia}, {Newberg}, {Couch}, \& {The
  Supernova Cosmology Project}}]{1999ApJ...517..565P}
{Perlmutter}, S., {et~al.} 1999, \apj, 517, 565

\bibitem[{{Phillips}(1993)}]{1993ApJ...413L.105P}
{Phillips}, M.~M. 1993, \apjl, 413, L105

\bibitem[{{Phillips}(2011)}]{2011arXiv1111.4463P}
---. 2011, ArXiv e-prints, arXiv:1111.4463

\bibitem[{{Phillips} {et~al.}(1999){Phillips}, {Lira}, {Suntzeff}, {Schommer},
  {Hamuy}, \& {Maza}}]{1999AJ....118.1766P}
{Phillips}, M.~M., {Lira}, P., {Suntzeff}, N.~B., {Schommer}, R.~A., {Hamuy},
  M., \& {Maza}, J. 1999, \aj, 118, 1766

\bibitem[{{Prieto} {et~al.}(2008){Prieto}, {Stanek}, \&
  {Beacom}}]{2008ApJ...673..999P}
{Prieto}, J.~L., {Stanek}, K.~Z., \& {Beacom}, J.~F. 2008, \apj, 673, 999

\bibitem[{{Riess} \& {Livio}(2006)}]{2006ApJ...648..884R}
{Riess}, A.~G., \& {Livio}, M. 2006, \apj, 648, 884

\bibitem[{{Riess} {et~al.}(1996){Riess}, {Press}, \&
  {Kirshner}}]{1996ApJ...473...88R}
{Riess}, A.~G., {Press}, W.~H., \& {Kirshner}, R.~P. 1996, \apj, 473, 88

\bibitem[{{Riess} {et~al.}(1998){Riess}, {Filippenko}, {Challis},
  {Clocchiatti}, {Diercks}, {Garnavich}, {Gilliland}, {Hogan}, {Jha},
  {Kirshner}, {Leibundgut}, {Phillips}, {Reiss}, {Schmidt}, {Schommer},
  {Smith}, {Spyromilio}, {Stubbs}, {Suntzeff}, \&
  {Tonry}}]{1998AJ....116.1009R}
{Riess}, A.~G., {et~al.} 1998, \aj, 116, 1009

\bibitem[{{Rodney} {et~al.}(2012){Rodney}, {Riess}, {Dahlen}, {Strolger},
  {Ferguson}, {Hjorth}, {Frederiksen}, {Weiner}, {Mobasher}, {Casertano},
  {Jones}, {Challis}, {Faber}, {Filippenko}, {Garnavich}, {Graur}, {Grogin},
  {Hayden}, {Jha}, {Kirshner}, {Kocevski}, {Koekemoer}, {McCully}, {Patel},
  {Rajan}, \& {Scarlata}}]{2012ApJ...746....5R}
{Rodney}, S.~A., {et~al.} 2012, \apj, 746, 5

\bibitem[{{Schlegel} {et~al.}(1998){Schlegel}, {Finkbeiner}, \&
  {Davis}}]{1998ApJ...500..525S}
{Schlegel}, D.~J., {Finkbeiner}, D.~P., \& {Davis}, M. 1998, \apj, 500, 525

\bibitem[{{Sullivan} {et~al.}(2006){Sullivan}, {Le Borgne}, {Pritchet},
  {Hodsman}, {Neill}, {Howell}, {Carlberg}, {Astier}, {Aubourg}, {Balam},
  {Basa}, {Conley}, {Fabbro}, {Fouchez}, {Guy}, {Hook}, {Pain},
  {Palanque-Delabrouille}, {Perrett}, {Regnault}, {Rich}, {Taillet}, {Baumont},
  {Bronder}, {Ellis}, {Filiol}, {Lusset}, {Perlmutter}, {Ripoche}, \&
  {Tao}}]{2006ApJ...648..868S}
{Sullivan}, M., {et~al.} 2006, \apj, 648, 868

\bibitem[{{Sullivan} {et~al.}(2010){Sullivan}, {Conley}, {Howell}, {Neill},
  {Astier}, {Balland}, {Basa}, {Carlberg}, {Fouchez}, {Guy}, {Hardin}, {Hook},
  {Pain}, {Palanque-Delabrouille}, {Perrett}, {Pritchet}, {Regnault}, {Rich},
  {Ruhlmann-Kleider}, {Baumont}, {Hsiao}, {Kronborg}, {Lidman}, {Perlmutter},
  \& {Walker}}]{2010MNRAS.406..782S}
---. 2010, \mnras, 406, 782

\bibitem[{{Sullivan} {et~al.}(2011){Sullivan}, {Guy}, {Conley}, {Regnault},
  {Astier}, {Balland}, {Basa}, {Carlberg}, {Fouchez}, {Hardin}, {Hook},
  {Howell}, {Pain}, {Palanque-Delabrouille}, {Perrett}, {Pritchet}, {Rich},
  {Ruhlmann-Kleider}, {Balam}, {Baumont}, {Ellis}, {Fabbro}, {Fakhouri},
  {Fourmanoit}, {Gonz{\'a}lez-Gait{\'a}n}, {Graham}, {Hudson}, {Hsiao},
  {Kronborg}, {Lidman}, {Mourao}, {Neill}, {Perlmutter}, {Ripoche}, {Suzuki},
  \& {Walker}}]{2011ApJ...737..102S}
---. 2011, \apj, 737, 102

\bibitem[{{Suzuki} {et~al.}(2012){Suzuki}, {Rubin}, {Lidman}, {Aldering},
  {Amanullah}, {Barbary}, {Barrientos}, {Botyanszki}, {Brodwin}, {Connolly},
  {Dawson}, {Dey}, {Doi}, {Donahue}, {Deustua}, {Eisenhardt}, {Ellingson},
  {Faccioli}, {Fadeyev}, {Fakhouri}, {Fruchter}, {Gilbank}, {Gladders},
  {Goldhaber}, {Gonzalez}, {Goobar}, {Gude}, {Hattori}, {Hoekstra}, {Hsiao},
  {Huang}, {Ihara}, {Jee}, {Johnston}, {Kashikawa}, {Koester}, {Konishi},
  {Kowalski}, {Linder}, {Lubin}, {Melbourne}, {Meyers}, {Morokuma}, {Munshi},
  {Mullis}, {Oda}, {Panagia}, {Perlmutter}, {Postman}, {Pritchard}, {Rhodes},
  {Ripoche}, {Rosati}, {Schlegel}, {Spadafora}, {Stanford}, {Stanishev},
  {Stern}, {Strovink}, {Takanashi}, {Tokita}, {Wagner}, {Wang}, {Yasuda},
  {Yee}, \& {Supernova Cosmology Project}}]{2012ApJ...746...85S}
{Suzuki}, N., {et~al.} 2012, \apj, 746, 85

\bibitem[{{Thomson} \& {Chary}(2011)}]{2011ApJ...731...72T}
{Thomson}, M.~G., \& {Chary}, R.~R. 2011, \apj, 731, 72

\bibitem[{{Tremonti} {et~al.}(2004){Tremonti}, {Heckman}, {Kauffmann},
  {Brinchmann}, {Charlot}, {White}, {Seibert}, {Peng}, {Schlegel}, {Uomoto},
  {Fukugita}, \& {Brinkmann}}]{2004ApJ...613..898T}
{Tremonti}, C.~A., {et~al.} 2004, \apj, 613, 898

\bibitem[{{Vernet} {et~al.}(2011){Vernet}, {Dekker}, {D'Odorico}, {Kaper},
  {Kjaergaard}, {Hammer}, {Randich}, {Zerbi}, {Groot}, {Hjorth}, {Guinouard},
  {Navarro}, {Adolfse}, {Albers}, {Amans}, {Andersen}, {Andersen}, {Binetruy},
  {Bristow}, {Castillo}, {Chemla}, {Christensen}, {Conconi}, {Conzelmann},
  {Dam}, {de Caprio}, {de Ugarte Postigo}, {Delabre}, {di Marcantonio},
  {Downing}, {Elswijk}, {Finger}, {Fischer}, {Flores}, {Fran{\c c}ois},
  {Goldoni}, {Guglielmi}, {Haigron}, {Hanenburg}, {Hendriks}, {Horrobin},
  {Horville}, {Jessen}, {Kerber}, {Kern}, {Kiekebusch}, {Kleszcz}, {Klougart},
  {Kragt}, {Larsen}, {Lizon}, {Lucuix}, {Mainieri}, {Manuputy}, {Martayan},
  {Mason}, {Mazzoleni}, {Michaelsen}, {Modigliani}, {Moehler}, {M{\o}ller},
  {Norup S{\o}rensen}, {N{\o}rregaard}, {P{\'e}roux}, {Patat}, {Pena}, {Pragt},
  {Reinero}, {Rigal}, {Riva}, {Roelfsema}, {Royer}, {Sacco}, {Santin},
  {Schoenmaker}, {Spano}, {Sweers}, {Ter Horst}, {Tintori}, {Tromp}, {van
  Dael}, {van der Vliet}, {Venema}, {Vidali}, {Vinther}, {Vola}, {Winters},
  {Wistisen}, {Wulterkens}, \& {Zacchei}}]{2011A&A...536A.105V}
{Vernet}, J., {et~al.} 2011, \aap, 536, A105

\bibitem[{{Wang} \& {Han}(2012)}]{2012NewAR..56..122W}
{Wang}, B., \& {Han}, Z. 2012, \nar, 56, 122

\bibitem[{{Whitaker} {et~al.}(2012){Whitaker}, {van Dokkum}, {Brammer}, \&
  {Franx}}]{2012ApJ...754L..29W}
{Whitaker}, K.~E., {van Dokkum}, P.~G., {Brammer}, G., \& {Franx}, M. 2012,
  \apjl, 754, L29

\bibitem[{{Wilson}(1939)}]{1939ApJ....90..634W}
{Wilson}, O.~C. 1939, \apj, 90, 634

\bibitem[{{Wuyts} {et~al.}(2011){Wuyts}, {F{\"o}rster Schreiber}, {van der
  Wel}, {Magnelli}, {Guo}, {Genzel}, {Lutz}, {Aussel}, {Barro}, {Berta},
  {Cava}, {Graci{\'a}-Carpio}, {Hathi}, {Huang}, {Kocevski}, {Koekemoer},
  {Lee}, {Le Floc'h}, {McGrath}, {Nordon}, {Popesso}, {Pozzi}, {Riguccini},
  {Rodighiero}, {Saintonge}, \& {Tacconi}}]{2011ApJ...742...96W}
{Wuyts}, S., {et~al.} 2011, \apj, 742, 96

\bibitem[{{Zwicky}(1938)}]{1938PhRv...53.1019Z}
{Zwicky}, F. 1938, Physical Review, 53, 1019

\end{thebibliography}
\bibliographystyle{apj}

\end{document}